\documentstyle[12pt,aasms4]{article}
\begin{document}

\title{Megamaser Disks in Active Galactic Nuclei}
\author{John F. Kartje and Arieh K\"onigl}
\affil{Department of Astronomy and Astrophysics and Enrico Fermi
Institute, University of Chicago, 5640 South Ellis Avenue, Chicago, IL 60637}
\affil{kartje@jets.uchicago.edu, arieh@jets.uchicago.edu}
\and
\author{Moshe Elitzur}
\affil{Department of Physics and Astronomy,
University of Kentucky, Lexington, KY 40506} 
\affil{moshe@pa.uky.edu}
\bigskip

\begin{abstract}
Recent spectroscopic and VLBI-imaging observations of bright
extragalactic H$_2$O maser sources have revealed that the
megamaser emission often originates in thin circumnuclear
disks near the centers of active galactic nuclei (AGNs). Using general
radiative and kinematic considerations and taking account of the
observed flux variability, we argue that the maser emission
regions are clumpy, a conclusion that is independent of the
detailed mechanism (X-ray heating, shocks, etc.) driving
the collisionally pumped masers. We examine scenarios in which the
clumps represent discrete gas condensations (i.e., clouds) and do not merely
correspond to velocity irregularities in the disk. We show that even two
clouds that overlap within the velocity coherence length along
the line of sight could account (through self-amplification) for
the entire maser flux of a high-velocity ``satellite'' feature
in sources like NGC 4258 and NGC 1068, and we suggest that cloud
self-amplification likely contributes also to the flux of the
background-amplifying ``systemic'' features in these
objects. Analogous interpretations have previously been proposed
for water maser sources in Galactic star-forming regions. We
argue that this picture provides a natural explanation
of the time-variability characteristics of extragalactic
megamaser sources and of their apparent association with Seyfert
2-like galaxies. We also show that the requisite cloud space densities and
internal densities are consistent with the typical values
of nuclear (broad emission-line region-type) clouds.

We examine two scenarios of clumpy disks in which the maser
emission is excited by a central continuum source. This
excitation mechanism was first considered in the context of
megamaser disks by Neufeld \& Maloney (1995), but our proposed
models are clearly distinct from their warped, homogeneous-disk
interpretation. In our first scenario we consider an annular disk
(or ``ring'') whose inner edge corresponds to the innermost radius of the
observed maser distribution and whose mass is dominated by the
clumped, high-density gas component. The shielding of the
high-energy continuum, which is
required in order that the gas remain molecular, can be provided in this
case by the dusty clouds themselves. We show that even the
simplest version of this model, in which the disk is flat and
the continuum radiation reaches the masing clouds through the
plane of the disk, can account for the maser observations in NGC
1068.  We point out the striking similarities between the
maser-ring properties as interpreted with this model and the
inferred characteristics of the circumnuclear ring (the
so-called CND) in the Galactic center, and we briefly discuss
the implications of such rings to the AGN accretion-disk paradigm.

Our second scenario is motivated by the apparent warps observed in
some of the imaged megamaser disks and by our finding that the
flat-disk version of the irradiated-ring scenario could apply to
a source like NGC 4258 only if the water abundace in the masing
clouds were higher than the value implied by equilibrium
photoionization-driven chemistry. This scenario is based on the
disk-driven hydromagnetic wind model originally proposed to
account for the molecular ``tori'' in Seyfert 2-like galaxies
and for several other observed phenomena in AGNs. In this
picture, the wind uplifts (by its ram pressure) and confines (by its
magnetic pressure) dense clouds fragmented from the disk, which
mase after they become exposed to the central radiation
field. Much of the requisite continuum shielding can be provided
in this case by the dusty portions of the wind. We show that
comparatively massive clouds that move in low-altitude, nearly circular orbits
could be shielded in this way, and we suggest that an apparent
warp in the maser distribution might arise under these
circumstances from an observational
selection effect induced by the strong vertical density
stratification that characterizes a centrifugally driven wind. We construct a
self-contained illustrative model where the wind transports the
bulk of the disk angular momentum, and we show that it is
consistent with the data for NGC 4258 as well as with the
advection-dominated accretion-flow interpretation of the
spectrum in this source.

\end{abstract}

% The different journals have different requirements for keywords.  The
% keywords.apj file, found on aas.org in the pubs/aastex-misc directory, 
% contains a list of keywords used with the ApJ and Letters.  These are 
% usually assigned by the editor, but authors may include them in their 
% manuscripts if they wish. 


\clearpage

\def\Tmas{\hbox{$\Theta_{\rm mas}$}}
\def\TT{\hbox{$\Theta^2_{\rm mas}$}}
\def\vlos{\hbox{$v_{\rm los}$}}
\def\T10{\hbox{$\Theta_{10}$}}
\def\Rc{\hbox{$R_{\rm c}$}}
\def\pc{\hbox{${\rm pc}$}}
\def\rin{\hbox{$r_{\rm in}$}}
\def\lc{\hbox{$\ell_{\rm c}$}}
\def\kms{\hbox{$\rm km \ s^{-1}$}}
\def\cmm{\hbox{$\rm cm^{-3}$}}
\def\nc{\hbox{$n_{\rm c}$}}

\section{Introduction}

Strong 22 GHz water maser emission has now been detected toward
the nuclei of some 20 active galaxies. These ``megamaser''
sources are generally associated with Seyfert 2 and LINER
galaxies whose nuclei are hidden by a fairly
large column of optically obscuring and X-ray--absorbing gas
(Braatz, Wilson, \& Henkel 1996, 1997). There is mounting
evidence, from spectral measurements and VLBI imaging, that in many
(if not all) of these objects at least part of the emission originates in
a rotating circumnuclear disk on scales of $\sim 0.1 - 1  \
\pc$. The best examples to date are
provided by NGC 4258 (e.g., Miyoshi et al. 1995; Greenhill et
al. 1995a) and NGC 1068 (e.g., Greenhill et al. 1996; Greenhill
\& Gwinn 1997), where the imaged maser spots are arranged in a thin and
apparently slightly warped disk. In both sources one detects
features that are located on the near side of the disk at the vicinity of its
apparent inner edge and have line-of-sight velocities that are
close to the systemic velocity of the galaxy, as well as high-velocity
``satellite'' features that are strung out along the projected
major axis (the midline) of the disk on either side of the nucleus (where
they give rise to red- and blue-shifted emission,
respectively). In NGC 4258 the systemic maser components lie near
$\rin \approx 0.13 \ \pc$ (assuming a
distance $D = 6.4 \ {\rm Mpc}$) and the high-velocity features
lie between $\sim 0.16$ and $\sim 0.26 \ \pc$; in NGC 1068
the inner edge is at $\rin \approx 0.65 \ \pc$
(for $D = 15 \ {\rm Mpc}$) and the radii of the high-velocity components
span the range $\sim 0.65-1.1 \ \pc$.
In the case of NGC 4258 the emission is dominated
by the low-velocity features, which are clustered
near the line of sight to the center (the centerline); in NGC 1068, the
redshifted high-velocity features dominate the measured flux, and the
emission from the inner edge is not confined to the vicinity of
the centerline (in fact, a full quarter of
the inner edge of the disk appears to be masing in this source, exhibiting a
linear increase of the line-of-sight velocity $v_{\rm los}$ with projected
distance from the center). The variation of the line-of-sight
velocity of the high-velocity features in NGC 4258 with
projected distance from the center indicates a Keplerian disk
around a central mass of $\sim 3.5 \times 10^7 M_{\odot}$;
the rotating-disk interpretation in this source is further supported by
measurements of the apparent centripetal acceleration of the
systemic masers (Haschick, Baan, \& Peng 1994; Greenhill et
al. 1995b). In NGC 1068 the $v_{\rm los}$ curve of the
high-velocity features exhibits a sub-Keplerian drop with
distance, and the inferred central mass is $\ga 10^7 \ M_{\odot}$.

Similar results have been obtained in several other sources,
although the data have often been more sketchy. In the case of NGC
4945 (Greenhill, Moran, \& Herrnstein 1997), red- and blue-shifted
maser features, symmetrically positioned with respect to the
systemic feature (at projected distances $\ga 0.2 \ \pc$) have been
imaged, and a central mass $\ga 10^6 \ M_{\odot}$ has been
inferred. A possible maser disk has also been
found in NGC 3079 (Trotter et al. 1998): in this case only
high-velocity (blue- and red-shifted) features, distributed over
$\sim 1.7 \ \pc$, have been
identified, and their motions are consistent with rotation
around a binding mass of $\sim 10^6 \ M_{\odot}$.  Other likely candidates for
an association of H$_2$O megamaser emission with a rotating disk
include NGC 2639 (Wilson, Braatz, \& Henkel 1995), the
Circinus galaxy (Greenhill et al. 1997a), and NGC 5793 (Hagiwara
et al. 1997). A noteworthy aspect of some of these apparent
disks is that they are misaligned with the local
symmetry axis. This is particularly pronounced in the case of NGC
1068, where the projected maser disk axis is tilted at an angle of $\sim
35^{\circ}$ with respect to the inner radio jet, which in turn is nearly
perpendicular to the plane of the circumnuclear gas distribution
(see Gallimore, Baum, \& O'Dea 1996, 1997; Tacconi et
al. 1997). A similar situation is indicated in NGC 3079,
where the apparent maser disk is tilted by $\sim
45^{\circ}$ with respect to the nuclear radio jet.

The imaged H$_2$O maser systems are
a unique diagnostic of the mass distribution
in the centers of AGNs. In particular, the high
mass concentrations inferred in sources like NGC 4258 currently
provide perhaps the strongest argument for the presence of
massive black holes in galactic nuclei (e.g., Maoz 1995,
1998; Rees 1998). By combining
centripetal-acceleration and proper-motion measurements of the systemic maser
spots, one can in principle also derive distance
ladder-independent values for the distances of these sources
(Herrnstein et al. 1997b). But the maser observations also
constitute a powerful tool for probing the nature of circumnuclear
disks in AGNs. In fact, since the bolometric luminosity in these
sources most likely derives from accretion, and since nuclear
jets have been detected in many of these objects, the inferred physical
conditions in the maser emission regions may have direct implications
to the study of accretion and outflow processes in AGNs.

There have been two distinct interpretations of the origin
of AGN megamaser disks, based on two alternative excitation
mechanisms. Although there is general agreement that the masers
are pumped collisionally, in one picture the maser emission
arises from the irradiation of the disk by the central continuum source (e.g.,
Neufeld, Maloney, \& Conger 1994), whereas in the alternative
interpretation the masers originate in shock waves that propagate
through the disk (e.g., Maoz \& McKee 1998, hereafter MM98). In the
disk-irradiation scenario it has been suggested that the disk
becomes exposed to the central radiation
field due to warping, so that the innermost radius $\rin$ from which
maser emission is observed effectively coincides with the onset
radius of the warp (Neufeld \& Maloney 1995). The location of
the latter is
in principle determined by the physical mechanism that
triggers the warping (e.g., radiation pressure-induced warping
instability [Maloney, Begelman, \& Pringle 1996] or an orbiting
massive object [Papaloizou, Terquem, \& Lin 1998]). In the shock
scenario, the high-velocity emission has been attributed to spiral shocks
arising from the steepening of self-sustaining waves that, in the
case of a non--self-gravitating, nearly Keplerian disk, are triggered
by external forcing (MM98). The trailing spiral
shocks envisioned in this model provide a natural explanation
of both the quasi-regular spacing of the high-velocity features
in sources like NGC 4258 and NGC 1068
and the tendency of the redshifted features to dominate the
blueshifted maser emission in most of the objects observed to date.

In the models considered so far in the literature the disk was
taken to be essentially homogeneous, and this assumption
was used in estimating the mass accretion rate through the maser
emission region. In the case of NGC 4258, Neufeld \& Maloney
(1995) considered an irradiated, viscous
accretion disk and concluded that a rest-mass conversion efficiency of
$\sim 0.25/\alpha$ (where $\alpha \le 1$ is the Shakura
\& Sunyaev [1973] viscosity parameter) was needed to account for the
bolometric luminosity $L_{\rm bol} \approx
10^{42} \ {\rm ergs} \ {\rm s}^{-1})$ of the central
source (Wilkes et al. 1995; Chary \& Becklin 1997). This
efficiency is a factor $\sim 250$ higher than the value inferred in this object
from an advection-dominated accretion flow (ADAF) interpretation
of its observed spectrum (Lasota et al. 1996; Chary \&
Becklin 1997; Herrnstein et al. 1998a). Although the latter
estimate applies to disk regions that are significantly smaller than the
maser emission zone, the two values
are mutually inconsistent if the disk has been
accreting in a steady manner over the last few million
years (Neufeld \& Maloney 1995). The ADAF-implied value also
appears to be incompatible with the radiation-induced warping
instability model for this galaxy (Maloney et al. 1996b). The
spiral-shock model, on the other hand, is evidently consistent
with the steady-state accretion rate ($\sim 2 \times 10^{-2} \,
\alpha \ M_{\odot} \ {\rm yr}^{-1}$; see Herrnstein et
al. 1998a) implied by the ADAF interpretation (MM98).

In this paper we take a closer look at the constraints imposed
by the observed maser emission on AGN disk models. In \S 2 we
argue, based on general radiative and
kinematic considerations as well as on the observed flux-variability
characteristics, that the maser emission regions are
clumpy. The likelihood of clumping, which is also
indicated by the spectroscopic and imaging data, has already been
recognized before (e.g., MM98), and could be
associated with either velocity irregularities or density inhomogeneities. 
The implications of a turbulent velocity field to the observed
maser spectra have been considered by Wallin, Watson, \& Wyld
(1998). Here we focus on the possibility that distinct gas
condensations (clouds) are a major (or even dominant) component
of the disk, and discuss the expected properties
of the resulting maser emission. The motivation for this study is
strengthened by our finding that the cloud parameters
implied by the maser observations are consistent with those
deduced by independent means but on roughly similar scales for optical/UV-
emitting clouds in Seyfert galaxies [the so-called broad
emission-line region (BELR) clouds].

In \S 3 we discuss the conditions for efficient water-maser
emission, starting with the general case of a collisionally
pumped gas and then specializing to irradiated clouds in
an AGN environment. Although the masing clumps could in
principle be excited in shocks as in the MM98 model, we
concentrate in this paper on scenarios
that are based on the alternative possibility of radiative
excitation. We construct two such models of irradiated clumpy disks
that are qualitatively different from the homogeneous-disk model
of Neufeld \& Maloney (1995). In the first such model (\S 4) we
consider a ``minimalist'' scenario in which the central
continuum radiation reaches the
clouds {\rm through the plane of the disk}: in this case there
is in principle no need to appeal to warping as in the homogeneous
irradiated-disk model. In this picture the megamaser disk
has a physical inner edge at $\rin$ and is therefore better described
as a ``ring'' (or ``annulus'').\footnote{A similar disk
geometry has been adopted also in the Maoz \& McKee
(1998) interpretation of NGC 4258.} Furthermore, because of the small
cloud volume-filling factor, the estimated disk mass is much
lower than that deduced from a homogeneous disk model. Focusing
on the case of NGC 1068, we discuss the striking similarity
between the maser ring as interpreted in this picture and the
circumnuclear dust ring (the so-called circumnuclear disk, or
CND) observed in the center of our own galaxy (e.g., Genzel,
Hollenbach, \& Townes 1994).

In applying the ``minimalist'' scenario considered in \S 4 to the
megamaser disk in NGC 4258 we find that a strictly flat disk
model is untenable unless the water abundance is significantly higher
than that implied by local photoionization-driven chemistry. 
Since the disk in this source appears to be warped (e.g.,
Herrnstein, Greenhill, \& Moran 1996), it is natural to consider a
scenario in which the maser distribution is not coplanar. As an
alternative to the warped-disk model of Neufeld \& Maloney
(1995) that specifically applies to a clumped mass distribution,
we present in \S 5 an interpretation of the high-velocity
features in terms of clouds that are uplifted from the surface of a
magnetized accretion disk by a centrifugally driven wind. Such
winds have previously been invoked to explain the origin of the
molecular ``tori'' in Seyfert 2-like galaxies and
other AGNs (K\"onigl \& Kartje 1994, hereafter KK94), and their
interaction with embedded
clouds may have additional important ramifications to active
galaxies (Kartje \& K\"onigl 1997, 1998). Using a self-contained
set of assumptions to represent a ``generic'' model of this type,
we show that this scenario is consistent with the maser
observations in a source like NGC 4258. We also demonstrate that
the previously noted discrepancies between the irradiated
accretion disk model for NGC 4258 and the ADAF interpretation of
its spectrum can in principle be resolved if a centrifugally driven wind
carries away a significant fraction of the disk angular momentum
in the maser emission region.

In \S 6 we summarize our results and discuss their implications
to the origin of maser emission and the nature of accretion
disks in AGNs.

\section{Maser Fluxes from AGN Disks}

\subsection{Flux Constraints}

Even with high-resolution VLBI imaging, the beam size is much larger than the
typical maser emission region in an AGN.
The flux measured from any small source is given by $F =
I\Omega$, where $I$ is the intensity and $\Omega$ is the solid
angle subtended by the source at the
observing position. If $A$ is the observed area of the source
and $D$ its distance, then
\begin{equation}
                  F = {2kT_{\rm b}\over \lambda^2}{A\over D^2} \, ,
\end{equation}
where $T_{\rm b}$ is the source brightness temperature and
$\lambda$ is the wavelength ($\approx 1.35 \ {\rm cm}$ for the water
maser transition). We now specialize to maser features (of arbitrary
shape) that lie in a disk whose plane is aligned with the observer. We denote
the half-length of the maser emission region along the line of sight by
$\ell$ and define the effective aspect ratio $a$ by
\begin{equation}
                  a = {\ell\over \sqrt{A/\pi}} \ .
\end{equation}
For $a$ larger than a few the maser is saturated and
its brightness temperature can be expressed as $T_{\rm b} \approx
10^{12} \, a_{10}^3 \ {\rm K}$, where $a_{10} = a/10$ (e.g.,
Elitzur, McKee, \& Hollenbach 1991, hereafter EMH91). The
observed maser flux is then 
\begin{equation}
F = 4.7 \times 10^{17} \ a_{10} \ \left ( {\ell \over D} \right )^2
\ {\rm Jy} \, .
\end{equation}
For a maser feature at a distance $r$ from the galactic center,
corresponding to an angular distance (in milliarcseconds) $\Tmas =
2.1 \times 10^8 \ (r/D)$, this can be rewritten in the form
 \begin{equation}
  F = 11 \ \TT \ a_{10} \ \left({\ell\over r}\right)^2 \  {\rm Jy} \, .
\end{equation}
Equations (3) and (4) are quite general, the only assumption
being that the pumping conditions are optimal. The pumping,
however, may be driven by any mechanism (shocks, X-ray heating,
etc.), and the geometry in the plane can be arbitrary.  The
expression (4) has the added convenience that, in VLBI-imaged
sources, $\Tmas$ is a measurable quantity.

One can derive an upper bound on the ratio $\ell/r$ in equation (4)
by using the velocity-coherence requirement for the maser
emission region. We assume Keplerian rotation and focus attention
on parcels of gas that give rise to a high-velocity maser feature. If one such
parcel is located on the disk midline at a distance $r=0.1 \
r_{0.1}$ pc from the center and has a line-of-sight velocity
$\vlos$, then another parcel that lies at a radius
$r + \delta r$ and is displaced from
the midline a distance $y$ will satisfy the coherence
requirement $\Delta \vlos \lesssim \Delta v/2$ (where
$\Delta v$ is the width of the maser line) so long as
\begin{equation}
   \left| \left(1 + {\delta r \over r} \right)^{-1/2}
          \left(1 + {y^2 \over (r + \delta r)^2} \right)^{-3/4} - 1 \right|
          \la {\Delta v \over 2 \, \vlos}  \, .
\end{equation}
This result implies that the emission region for any given maser
feature is confined to
a velocity-coherence ``box'' of half-width (along the midline) 
\begin{equation}
w_{\rm coh}  \approx \left ( \frac{\Delta v}{\vlos}
\right ) \ r \approx 3.1 \times 10^{14} \ r_{0.1} \
\left ( \frac{\Delta v}{10^{-3}\ \vlos} \right ) \ {\rm cm}
\end{equation}
and half-length (along the line of sight)
\begin{equation}
\ell_{\rm coh} \approx \left ( \frac{2 \Delta v}{3 \vlos}
\right )^{1/2} \ r \approx 8.0 \times 10^{15} \ r_{0.1} \
\left ( \frac{\Delta v}{10^{-3} \ \vlos} \right )^{1/2} \ {\rm cm}
\end{equation}
(cf. MM98).\footnote{The height of the
velocity-coherence ``box'' is determined by such factors as the vertical
acceleration of the emitting gas, the thickness of the layer in
which optimal masing conditions prevail, and the telescope
beamwidth. In the models discussed in this paper, the
relevant factor typically turns out to be the beamwidth
($\la 0.5 \ {\rm mas}$).} Equations (5)--(7) again have a
rather general validity and, in fact, apply to any
masing gas that is distributed in a Keplerian disk.

Equations (4) and (7) can be combined to give
 \begin{equation}
  F \la 0.7\ a_{10} \ {\T10}^2 \left ( {\Delta v\over
  10^{-3}\ \vlos} \right ) \  {\rm Jy} \, ,
 \end{equation}
where $\T10 = \Tmas/10$. The
maximum flux from a maser feature would be produced if the
emission came from the entire velocity-coherence ``box.''
Adopting a slab-like geometry for this ``box,'' the observed
maser emission area is $A \approx 2 w_{\rm coh}^2$ (Elitzur,
Hollenbach, \& McKee 1992, hereafter EHM92), and hence $a_{10} \approx 0.125 \
\ell_{\rm coh}/w_{\rm coh}$. Substituting from equations (6) and (7), one gets 
$a_{10} \approx 3.2 \, ({\Delta v/ 10^{-3} \, \vlos})^{-1/2}$, so
equation (8) becomes 
\begin{equation}
F_{\rm max} \approx 2.4 \ \T10^2 \left ({\Delta v\over
  10^{-3}\ \vlos} \right )^{1/2} \  {\rm Jy} \, .
\end{equation}

Specializing to NGC 1068, for which we adopt as fiducial mean
parameters $\Tmas \approx 10.9$ and $(\Delta v/v_{\rm los})
\approx 3.5 \times 10^{-3}$, we infer from equation (9) that
$F_{\rm max} \approx 5.3 \ {\rm Jy}$. This is about an order of
magnitude larger than the maximum flux ($\sim 0.4 \ {\rm Jy}$)
reported for the high-velocity features in this object (e.g.,
Greenhill et al. 1996). The implication of this
result is that only a fraction of the potentially observable volume in this
source actually contributes to the maser emission. In the case
of NGC 4258 the measured maximum flux in the high-velocity
features (e.g., Moran et al. 1995) sometimes approaches the estimated value of
$F_{\rm max}$ [$\sim 1.1 \ {\rm Jy}$, using $\Tmas \approx 6.7$ and
$(\Delta v/v_{\rm los}) \approx 10^{-3}$]; however, in this
case, too, most of the high-velocity features typically have lower
fluxes. There is, in fact, an independent piece of evidence that
indicates that the high-velocity maser emission zone in NGC
4258 does not comprise the entire velocity-coherent region.
From equation (3)
it is seen that $F \propto a_{10} \ \ell^2$. Taking $a_{10}
\propto \ell_{\rm coh}/w_{\rm coh}$, using equations (6) and (7), and noting
that, for a Keplerian velocity field, $r \propto \vlos^{-2}$, one
obtains $F \propto \vlos^{-4.5}$. If, instead, one assumes that
$a_{10}$ is more-or-less uniform throughout the emission region (as
would be the case, for example, for a collection of similar
X-ray--irradiated clouds), then an even steeper dependence, $F
\propto \vlos^{-5}$, is obtained. Now, in the first
interferometric mapping on 1994 April 26, the emission from the
high-velocity features in NGC 4258 did exhibit a trend of $F$ decreasing with
$\vlos$ (see Fig. 1 in Miyoshi et al. 1995), but the decline was much
milder than $\vlos^{-4.5}$ (see also Fig. 11 in Nakai et
al. 1995). In mappings performed on 1995 January 8 and 1995 May 28
the highest-velocity peaks rose sharply, so the flux near
the high end of the spectrum was actually {\em increasing}
with $\vlos$ (Herrnstein et al. 1998b),
the opposite of the trend predicted by velocity coherence. Hence, independent
of the normalization of the pump estimates, one concludes that
the line-of-sight maser sizes do not sample the source velocity
field and therefore are not directly determined by the disk dimensions.

The inference that the size of the emission region in
objects like NGC 1068 and NGC 4258 is
independent of that of the velocity-coherence ``box'' is
consistent with the MM98 model, in which the dimensions of the
masing region are
determined by the spiral-shock characteristics. One can, however,
argue, based on measurements of flux variability, that the
emission must arise in distinct clumps (representing velocity or
density irregularities within the disk) even if the large-scale
structure of the masing region is defined by shocks. In the case
of NGC 4258, VLBA observations reported by Herrnstein et
al. (1998b) have revealed strong (up to a factor of $\sim 3$)
variations in the flux measured from some of the high-velocity
features on time scales $\Delta t_{\rm var} \la 8\ \Delta t_8$
months. A similar behavior was also noted in spectroscopic observations of
NGC 1068 (Nakai et al. 1995) and NGC 5793 (Hagiwara et
al. 1997). Such variability (which is manifested by both
increases and decreases in the maser flux) is not
expected in a uniform emission region, but it can be readily
interpreted in terms of the poloidal ($r-z$) motion (at a speed
$\ga R_{\rm c}/\Delta t_{\rm var} \approx 5 \ R_{\rm c\, 13} \ \Delta
t_8^{-1} \ {\rm km} \ {\rm s}^{-1}$) of distinct
masing clumps of transverse size $R_{\rm c} = 10^{13} \, R_{\rm c\, 13} \
{\rm cm}$ that move into (or out of) alignment with
each other along the line of sight to the observer (see \S
2.2). In order for this behavior to be observed, the clump
overlap probability must be high (see eq. [17] below), which is
equivalent to the cloud covering factor in the
emission region being large. This, in turn, explains why the
flux from a given feature can occasionally decline without the
entire maser emission region fading away: when one pair of clouds drifts out of
alignment, another pair is likely to drift into alignment on a
similar time scale.

The presence of clumps has also been deduced in sources like NGC
4258 from spectral and imaging observations of the systemic
masers (e.g., Haschick et al. 1994; Greenhill et al. 1995a,b;
Herrnstein et al. 1997b). It is interesting to note in this
connection that the latter features also exhibit strong
variability (including a factor of $\sim 10$ flares lasting on the
order of one month; Greenhill et al. 1997b) that could
potentially arise from similar clump-alignment effects (see \S
2.2), although variability in the central radio-continuum source 
may also be relevant in this case (e.g., Herrnstein et al. 1997a).

\subsection{Flux from Masing Clouds}

We now consider the maser emission from individual clouds that
have a Keplerian disk-like distribution. We start with the case
of saturated masers and then address the
unsaturated ones at the end of this subsection. As discussed,
for example, by EHM92, strongly amplified masers have a high degree of
beaming and therefore have a filamentary appearance. This is
true even for spherical masers, whose observed transverse sizes
are significantly smaller than their radii on account of
amplification-bounded beaming. Cylindrical masers, on the other
hand, are matter bounded, so their observed radii are the same
as their projected physical sizes. For the sake of simplicity we
consider in what follows cylindrical masers of radius $\Rc$ and
half-length $\lc = a \, \Rc$, although it should be borne in mind
that the actual physical dimensions of the masing clouds could
be characterized by smaller aspect ratios. Real clouds in a
differentially rotating disk would probably
be elongated at least to some degree in the azimuthal direction and hence
may be better represented by cylinders than by spheres.

Focusing on the high-velocity maser features, which we represent
as filaments viewed end on, the flux from an individual cloud is
given by
\begin{equation}
F_{\rm c} \approx 5.0 \times 10^{-6} \ D_{10}^{-2} \ R_{\rm c\, 13}^2 \
a_{10}^{3} \ {\rm Jy}
\end{equation}
(see eq. [3]), where we set $D_{10} = (D/10\, {\rm Mpc})$.
This value is much smaller than the
flux ($F \ga 0.1 \ {\rm Jy}$) typically measured from the bright
high-velocity maser features in AGNs. However, if two cloud
emission regions that lie within the velocity-coherence ``box''
overlap along the line of sight, the flux from the background
cloud can be strongly amplified by the foreground emission
region (Deguchi \& Watson 1989; EMH91). The effective brightness
temperature $T_{\rm b, \, eff}$ for two
overlapping masers that are separated by a distance $s$, each
characterized by a brightness temperature $T_{\rm b}$ and
aspect ratio $a$, is 
\begin{equation}
T_{\rm b, \, eff} \approx  \frac{16}{11} \left ( \frac{s}{a\,
\Rc} \right )^2 \left [ 1 + \left ( \frac{s}{d_{\rm i}} \right
)^2 \right ]^{-1} \, T_{\rm b} \, ,
\end{equation}
where
\begin{equation}
d_{\rm i} = 10^{18} a_{10}^2 \ \gamma_5^{1/2} \ R_{\rm c\, 13} \ {\rm cm}
\end{equation}
is the interaction distance, expressed in terms of the ratio
$\gamma = 10^5 \, \gamma_5$ of the maser saturation intensity to the
unsaturated source function (EMH91; see also Elitzur 1992). The
corresponding flux is given by
\begin{equation}
F \approx 7.3 \times 10^{-2} a_{10} \ D_{10}^{-2} \
s_{16}^2 \ (1 + 10^{-4} \, s_{16}^2 \, d_{\rm i\, 18}^{-2})^{-1}
\ {\rm Jy} \, ,
\end{equation}
where we put $ s_{16} = (s/10^{16}\, {\rm cm})$ and $d_{\rm i\,
18} = (d_{\rm i}/10^{18}\, {\rm cm})$. Note the
interesting result that, in the limit $s \ll d_{\rm i}$, the
self-amplified flux is independent of the cloud size (depending on
the cloud dimensions only through the aspect ratio $a$) and is determined
primarily by the line-of-sight separation between the clouds ($F
\propto s^2$). It is seen that even two such clouds could in
principle account for the entire measured flux $F =
0.1 \ F_{0.1} \ {\rm Jy}$ of a maser feature if they were
separated by a distance greater than
\begin{equation}
s_{\rm min} \approx
1.4 \times 10^{16} \ F_{0.1}^{1/2} \ D_{10} \ 
a_{10}^{-1/2} \ {\rm cm}
\end{equation}
along the line of sight (expression valid for $s_{\rm min} \ll
d_{\rm i}$). The value of $s_{\rm min}$ would be larger if
the cloud emission regions overlapped only partially. On the
other hand, the value of $s_{\rm min}$ could be smaller if more
than one interacting pair of masing clouds contributed to the observed flux.

For self-consistency, $s_{\rm min}$ is required to be $< 2 \ \ell_{\rm
coh}$ (eq. [7]). This condition appears to be satisfied in NGC
4258, where (using $D_{10} \approx 0.64$, $a_{10} \approx
1$, and $F_{0.1} \approx 10$ in eq. [14] and evaluating eq. [7]
at $r = 0.21 \ \pc$), $s_{\rm min}/\ell_{\rm coh} \approx
1.4$, as well as in NGC 1068, where (with $D_{10} \approx 1.5$, $a_{10} \approx
1$, $F_{0.1} \approx 4$, and $r = 0.88 \ \pc$) $s_{\rm
min}/\ell_{\rm coh} \approx 0.27$. The fact that
the estimated value of $s_{\rm min}$ turns out not to be much
smaller than the maximum value of $\ell_{\rm coh}$ 
suggests that there is a strong observational bias for detecting
high-velocity megamaser features in Seyfert 2-like galaxies. In
these AGNs the disk
rotation axis is nearly perpendicular to the line of sight and the
corresponding value of $\ell_{\rm coh}$ is large: by contrast,
AGNs with disks that are viewed at smaller angles to the axis have a
comparatively low probability of overlap for masing clouds that
satisfy the velocity-coherence constraints. Note that, if
cloud ``stretching'' is tied to the velocity shear in the disk,
then a large inclination angle would also be favored by the dependence
of the flux on the filament aspect ratio $a$.

One can show that masing clouds with typical parameters that are
even moderately elongated are likely to be saturated.
The saturation condition for a cylindrical maser can be written as
\begin{equation}
\frac{{\rm exp} \ [a \, \kappa_0 \, \Rc ]}{a} = 4 \, \sqrt{\gamma}\ ,
\end{equation}
where $\kappa_0$ is the unsaturated line-center absorption coefficient
(EMH91). When all parameters except for $a$ are fixed, this
relation determines $a_{\rm sat}$, the minimum aspect ratio for saturation.
As an illustration, when the masing gas is characterized by a hydrogen number
density of $10^9 \ \cmm$, temperature of $400 \ {\rm K}$, maser line
width of $1 \ \kms$, and fractional water abundance of
$10^{-5}$, then $\kappa_0 \, \Rc = 0.8,\ 1.2,\ 2.7$,
and 3.7 for $R_{\rm c\, 13}=0.5,\ 1,\ 5,$ and 10, respectively. The
solutions of equation (15) are then $a_{\rm sat}
\approx 12.9,\ 8.0,\ 3.0,$ and 2.2, respectively, where we set
$\gamma_5 = 1$. The corresponding values for a fractional water
abundance of $10^{-4}$ are $\kappa_0 \, \Rc = 2.7, \ 3.7,\ 6.1$,
and 7.2, and $a_{\rm sat} \approx 3.0,\ 2.2,\ 1.2$, and 1.0, respectively.
[Note that, in order to be considered
filamentary, the maser must satisfy $a \gg {\rm max} \,
\{1,\kappa_0 \Rc/4\}$ (EMH91).] For comparison, a
spherical maser of radius $\Rc$ saturates when $\kappa_0 \, \Rc
\approx 7.5$ (assuming again $\gamma_5 = 1$), at which point it
appears as a cylinder with an effective aspect ratio of $\sim 2.9$ (see EHM92).
These results indicate that masing clouds that are at least
somewhat elongated and have $\kappa_0 \, \Rc$
values of the order of a few would be saturated.

In a source like NGC 4258 one can use information on the
systemic masers to help assess whether the saturation condition
on the high-velocity features is satisfied. Based on the location of
the systemic masers relative to the nuclear radio jet and on the
apparent correlation between the jet continuum and maser flux
densities, it has been inferred that the systemic maser features in this
object act as amplifiers of the background radio continuum
(e.g., Herrnstein et al. 1997a). On the assumption that each
such feature is produced by the same type of filamentary clouds that give
rise to the high-velocity emission (except that they are now
viewed at a right angle to their axes), the brightness temperature
of the systemic masers is found to be
$T_{\rm b,\, sys} \approx 2.6 \times 10^{15} \, (F/2 \, {\rm Jy}) \
(a/2)^{-1} \ (\Rc/5 \times 10^{13} \, {\rm cm})^{-2} \ {\rm
K}$. Setting $T_{\rm b,\, sys} =
{\rm exp}\ (\tau_{\rm sys,\, tot}) \, T_{\rm b,\, cont}$, where
$T_{\rm b,\, cont}$ ($\la 10^8 \ {\rm K}$; Greenhill et
al. 1995a; Herrnstein et al. 1998a) is the brightness
temperature of the amplified background continuum, we deduce that 
the total optical depth $\tau_{\rm sys,\, tot}$ of the systemic masers
is $\sim 17.1$. If the maximum systemic maser flux is produced by
two fully overlapping clouds then it follows that $\kappa_0 \, \Rc \approx
\tau_{\rm sys,\, tot}/4 \approx 4.27$. Now, the total optical depth
$\tau_{\rm hi,\, tot}$ for the high-velocity masers is given in
terms of the maser brightness temperature $T_{\rm b,\, hi}
\approx 1.6 \times 10^{15} \, (F/0.5 \, {\rm Jy}) \
(\Rc/5 \times 10^{13} \, {\rm cm})^{-2} \ {\rm K}$ and the
excitation temperature $T_{\rm x} \approx 10^2 \ {\rm K}$ (e.g.,
Elitzur 1992) by $\tau_{\rm hi,\, tot} \approx {\rm ln} \ (T_{\rm
b,\, hi}/T_{\rm x})$. Attributing the high-velocity emission to
two overlapping filaments viewed end on, we obtain for the above
fiducial values $a \, \kappa_0 \, \Rc \approx 7.6$, which, when
combined with the systemic-masers result, implies $a \approx
1.78$. This value is just shy of the saturation aspect ratio
$a_{\rm sat} \approx 1.81$ obtained from equation (15) for
$\kappa_0 \, \Rc = 4.27$. 

From the above estimate we conclude that the saturation
assumption for the high-velocity masers in a source like NGC
4258 is probably valid, although it might be only marginally so
in clouds with large values of $\Rc$. (Recall in
this connection that an upper bound on $\Rc$ is provided by
the half-width $w_{\rm coh}$ of the
velocity-coherence ``box,'' given by eq. [6].) Our
interpretation of the high-velocity features in terms of two
overlapping filaments would not fundamentally change even if the
filaments were unsaturated. In that case the brightness
temperature of each individual filament would drop from $\sim
10^{12} \, a_{10}^3 \ {\rm K}$ to $\sim 10^2 \, {\rm exp} \ (2\,
a\, \kappa_0 \, \Rc) \ {\rm K}$, but the self-amplification
factor would increase from the value given by equation (11) to
$\sim {\rm exp} \ (2\, a\, \kappa_0\, \Rc)$, resulting in the
same final brightness temperature $T_{\rm b,\, hi}$. [Note that
the two masers would remain unsaturated only if the condition
$(\pi \, \Rc^2/4 \, \pi \, s^2) \ {\rm exp} \ (2\, a\,
\kappa_0\, \Rc) > 1$ is satisfied, which would be the case only
if the cloud separation $s$ is not too large.] Even though the
amplification factor in the unsaturated case no longer depends
on $s$ (in contrast with eq. [11]), the argument that
high-velocity maser features should be seen only at large
inclination angles to the disk rotation axis still applies since
a comparatively large value of $\ell_{\rm coh}$ is still
required to attain a measurable probability of cloud
overlap.

If the systemic maser emission in a source like NGC 4258 arises
in two or more overlapping clumps that amplify the background
continuum, then their detection probability would also be
maximized at large inclination angles. The coherence half-length
near the centerline of a Keplerian disk viewed along its plane can exceed
the midline value (eq. [7]) if the emission originates
sufficiently close to the centerline. Exactly along the
centerline all points have the same line-of-sight speed
($\vlos = 0$), whereas for a masing parcel at a radius $r$ where
$\Delta v \la \vlos \la v_{\rm K}/2$ (with $v_{\rm K}$ denoting the
Keplerian rotation speed) the coherence half-length is $\sim (\Delta v/3\,
\vlos)\, r$ (which differs slightly from the expression in MM98);
these values may be reduced if the disk plane is
tilted with respect to the line of sight. Since the coherence length
is maximized along the centerline, one would expect the systemic
maser emission to peak there: in fact, the maser spectrum in NGC
4258 exhibits a dip at $\vlos = 0$, which is most naturally
attributed to absorption outside the emission region (Watson \&
Walin 1994; MM98; see \S 4). The systemic maser
emission is thus maximized on either side of the centerline,
where the velocity coherence length along the line-of-sight from
the emission point does not reach all the way into the absorption
region. For these features the line-of-sight coherence length
may still be comparable to, or even exceed, the value associated
with the high-velocity masers. Hence, if the high-velocity
maser emission originates in overlapping clouds, at least part
of the systemic maser emission might too. The motion of
individual clouds into or out of alignment along the line of
sight (see \S 2.1) could then explain also some of the strong flux
variability exhibited by the systemic masers (e.g., Greenhill
et al. 1997b). We address additional aspects of the systemic
masers in \S 4.

\section{Masing Conditions}

Optimal emission in the $6_{16} - 5_{23} \ 22 \ {\rm GHz}$ maser
transition from collisionally pumped water molecules
is achieved for hydrogen-nuclei densities $n_{\rm H} \sim 10^8 - 10^9 \ 
{\rm cm^{-3}}$, temperatures $T \ga 250\ {\rm K}$, and
fractional water abundances $x({\rm H_2O}) \equiv n({\rm
H_2O})/n_{\rm H} \sim 10^{-5} - 10^{-4}$ (e.g., Elitzur
1992). We note in passing that the required densities fall
within the inferred characteristic range in BELR clouds (e.g.,
Baldwin et al. 1995). Neufeld et al. (1994) showed that
these conditions are attained in a dusty gas of density
$n_{\rm H} \ga 10^8 \ {\rm cm^{-3}}$ that is irradiated by a
typical AGN X-ray spectrum if the
column density $N_{\rm H}$ exceeds $\sim 10^{23} \ {\rm
cm^{-2}}$ (see also Pier \& Voit 1995; Wallin \& Watson 1997).
The thermal structure of a photoionized, optically thin medium
is in general determined by the ionization parameter $\Xi \equiv
F_{\rm ion}/p \, c$, where $F_{\rm ion}$ is the local ionizing
flux and $p = n_{\rm H} \, k\, T$ (with $k$ being Boltzmann's
constant) is the gas thermal pressure (modulo a molecular-weight
factor). In the case of a dusty,
molecular gas, which absorbs the low-energy ionizing photons up
to an energy that depends on the magnitude of the column
density, one can construct an effective ionization parameter
that explicitly accounts for the attenuation of the incident
ionizing flux $F_{\rm ion,0}$ and that can be approximated by
$\Xi_{\rm eff} \approx F_{\rm ion,0}/N_{\rm H}\, p \, c$
(Maloney, Hollenbach, \& Tielens 1996). Neufeld et al. (1994)
identified the maximum value of $\Xi_{\rm eff}$ where a stable
molecular phase first appears, and Neufeld \& Maloney (1995), in
turn, used this value to obtain the minimum
pressure that is required (given $F_{\rm ion,0}$ and $N_{\rm H}$) for the
gas in an irradiated disk to be molecular. In the models
considered in this paper it is more convenient to inquire
about the minimum shielding column, $N_{\rm min}$, that must be
present for given irradiating flux and gas pressure in order for
the gas to be molecular. To derive the gas thermal properties in our
particular applications we employ the photoionization code {\em
CLOUDY} (Ferland 1996) and compute the run of
temperature and water abundance as a function of depth into a
slab of given (uniform) density that is irradiated by an
external continuum flux and is shielded by a dusty gas of column
density $N_{\rm shield}$. The grain and elemental abundances
are taken to be those of the local ISM. Figure 1 shows the run
of $N_{\rm min}$ obtained in this way as a function of $p/k$
for clouds (of hydrogen-nuclei density $n_{\rm c}$ and temperature
$T_{\rm c}$) that are located in the H$_2$O maser emission
regions in NGC 1068 and NGC 4258. For NGC 1068 we adopt
the spectrum given by Pier et al. (1994) with $L_{\rm bol} \approx
4 \times 10^{44} \ {\rm ergs} \ {\rm s}^{-1}$, and set $r = 1 \
{\rm pc}$, whereas for NGC 4258 we use the spectral
information from Wilkes et al. (1995) and Chary \& Becklin
(1997) and set $L_{\rm bol} \approx 10^{42} \ {\rm ergs} \ {\rm
s}^{-1}$ and $r = 0.25 \ {\rm pc}$. Approximate polynomial fits
to the function $N_{\rm min}(p)$ for these two
cases are given in the caption to Figure 1. Since $T_{\rm c}$ is
determined by the value of the ionization parameter, $N_{\rm
min}$ can be regarded as being just a function of $\nc$.

Molecular gas in an AGN can remain dusty beyond the dust
sublimation radius $r_{\rm sub}$, defined as the distance
interior to which dust grains exposed to the ambient radiation
field are evaporated. The primary significance of dust to AGN
maser emission is that, by absorbing UV and soft X-ray photons
from the central continuum, it lowers the ionization state and
hence the temperature of the gas so that water molecules can
survive (or form) at distances from the center where the gas
would otherwise be hot and atomic. 
In the presence of a sufficiently large shielding column the
temperature drops below $T \approx 1500 \ {\rm K}$ and the gas
remains (or becomes) molecular (see, e.g., Fig. 1 in Neufeld et
al. 1994). At this temperature ${\rm H_2}$ is 
efficiently produced by the reaction ${\rm H^- + H \rightarrow H_2} + e$,
which permits water formation via the reaction network ${\rm O +
H_2}  \rightarrow  {\rm OH + H}$; ${\rm OH + H_2} \rightarrow {\rm H_2O + H}$.
Therefore, even if the water abundance in the disk were
initially low, it could become comparatively high [$x({\rm
H_2O}) \ga 10^{-5}$] after it gets exposed to the (shielded)
central continuum. The magnitude of the sublimation radius depends on the
detailed spectral energy distribution and scales with the source
bolometric luminosity as $L_{\rm bol}^{1/2}$. Its precise value also
depends on the grain composition and size distribution (e.g.,
Laor \& Draine 1993). Using the same prescription as
in KK94 (see their Appendix B), we infer $r_{\rm sub} \approx 0.23 \ (L_{\rm
bol}/4 \times 10^{44} \ {\rm ergs \ s^{-1}})^{1/2} \ {\rm pc}$
in NGC 1068 and $r_{\rm sub} \approx 0.015 \ (L_{\rm
bol}/10^{42} \ {\rm ergs \ s^{-1}})^{1/2} \ {\rm pc}$
in NGC 4258. In a similar manner we deduce $r_{\rm sub}
\approx 0.15 \ (L_{\rm bol}/10^{44} \ {\rm ergs \ s^{-1}})^{1/2} \ {\rm pc}$
in NGC 4945 (using the spectral information in
Iwasawa et al. 1993) and $r_{\rm sub} \approx 0.25 \ (L_{\rm
bol}/3 \times 10^{44} \ {\rm ergs \ s^{-1}})^{1/2} \ {\rm pc}$
in NGC 3079 (using the scaled nuclear bolometric luminosity
from Hawarden et al. 1995 together with the inferred spectral
shape of NGC 4258). If the maser
emission is excited by continuum irradiation then $r_{\rm sub}$
should not exceed the inner radius $\rin$ of the maser emission
region: the observed values of $\rin$ (see \S 1) evidently satisfy this
inequality for the above sources.\footnote{The
possible relevance of the dust
sublimation radius to the location of the inner boundary of the
maser emission region was already pointed out by Greenhill et
al. (1996) in their discussion of the NGC 1068 VLBI imaging data.}

When the column density of dusty gas that shields the continuum
source becomes large enough, the cloud temperature remains
below $250 \ {\rm K}$ and the water level-population inversion required for
maser emission does not occur. We propose that, in the
context of clumpy disk models, this condition (rather than the
transition from molecular to atomic gas invoked in the
homogeneous-disk model of Neufeld \& Maloney 1995) determines
the outermost radius $r_{\rm out}$ of the maser emission
region. Specifically, for a given incident continuum spectrum, the
condition $T \approx 250 \ {\rm K}$ at the upstream edge of a cloud
determines the magnitude of the intervening column density,
which we denote by $N_{\rm max}$. For a given gas distribution,
this value, in turn, fixes $r_{\rm out}$. Since the temperature
is tied to the value of the effective ionization parameter
$\Xi_{\rm eff}$, which depends on the gas density, $N_{\rm max}$
is a function of $n_{\rm c}$ (and, like $N_{\rm min}$, scales
roughly as $\nc^{-1}$).

Clouds located at radii $r < r_{\rm out}$ have temperatures
$T > 250 \ {\rm K}$, so their water molecules are collisionally 
pumped. Adopting again the filamentary geometry introduced in \S
2.1, the maser efficiency in these clouds can be parametrized in
terms of the effective emission measure
\begin{equation}
\xi = 0.2\, x_{-5}({\rm H_2O})\, n_{\rm c\, 9}^2 \, R_{\rm c\,
13}/\Delta v_5 \, ,
\end{equation}
where $x_{-5}({\rm H_2O})=x({\rm H_2O})/10^{-5}$, $n_{\rm c\,
9} = (n_{\rm c}/10^9 \, {\rm cm^{-3}})$, and
$\Delta v_5 =(\Delta v / 10^5  \, {\rm cm \ s^{-1}})$ (Elitzur,
Hollenbach, \& McKee 1989, hereafter EHM89).
Optimal masing conditions obtain for $1 \la \xi \la 100$.  For
$\xi \ga 100$, the (dusty) cloud column density $\sim
10^{23} \, \xi \, \Delta v_5 /(x_{-5} \, n_{\rm c\, 9})\ {\rm
cm}^{-2}$ becomes too large
for $T$ to remain above $\sim 250$ K.\footnote{In the case of a
dust-free gas, a similar upper limit on the optical depth is commonly
attributed to the fact that, for $\xi \ga 100$,
the maser quenches due to infrared photon trapping that tends to
thermalize the level populations (EHM89). However, in the case
of a dusty gas, Collison \& Watson (1995; see also Wallin \&
Watson 1997) have shown that, if the grain and gas temperatures
differ by more than $\sim 10$ K, the pumping is no longer
quenched but instead remains constant with increasing optical
depth. Although this condition is generally satisfied
for the clouds that we model in this paper, $\xi \approx 100$ is
still the relevant upper limit because of the temperature
constraint mentioned in the text. In any case, the clouds in the
scenarios that we explore are sufficiently small that their
individual column densities do not even approach this upper limit.}

\section{Masing Rings}

In this section we incorporate the masing conditions derived
in \S 3 into a set of self-consistency constraints for a
``minimalist'' model of an irradiated disk, and we demonstrate
that this model can account for the H$_2$O maser observations in a
source like NGC 1068. A schematic
illustration of the envisioned scenario is given in Figure 2. We
consider a planar ring of inner radius $\rin$ that surrounds the
central continuum
source. The ring is composed of dusty molecular clumps and
possibly also of a dusty interclump medium, but the total column
density through the disk is assumed to be sufficiently low that
the ionizing radiation can reach the masing clouds {\em through
the plane of the ring}. For simplicity, we consider a collection
of identical cylindrical clouds of density $n_{\rm c}$, radius
$\Rc$, and aspect ratio $a \approx 10$, whose axes are roughly
perpendicular to the radius vectors from the center.\footnote{A
more physical characterization of the clouds would have been in
terms of their pressure $p$ rather than $\nc$ (see Neufeld et
al. 1994), but for the narrow range in temperatures that is of
interest to us the distinction is not significant.} We further
assume that the clouds are distributed uniformly throughout the
disk. Focusing attention on the high-velocity features, which we interpret in
terms of overlapping clouds (see \S 2), we derive an
approximate estimate of the minimum number density of clouds
${\cal N}_{\rm c,\, min}$ that is required for our
line of sight through the disk plane to intercept, on the
average, two overlapping clouds within a velocity coherence
length, by setting ${\cal N}_{\rm c,\, min} \approx 1/\pi \,
\Rc^2 \, \ell_{\rm coh}$. Substituting from equation (7), we obtain
\begin{equation}
{\cal N}_{\rm c,\, min} \approx 1.2 \times 10^{13} \
(\Delta v/10^{-3} \ v_{\rm los})^{-1/2} \ R_{\rm c\, 13}^{-2} \
r_{0.1}^{-1} \ \ {\rm pc}^{-3} \, .
\end{equation}
It is noteworthy that this fiducial value of the cloud space
density, which is applicable to any scenario that interprets the
high-velocity maser features in terms of overlapping clumps, is
consistent with independent
estimates for BELR clouds in typical AGNs (e.g., Arav et al. 1997). 

The minimum cloud space density given by equation (17)
corresponds to a cloud volume filling factor
\begin{equation}
f_{\rm V,\, min} \approx (2\, \Rc/a\, \ell_{\rm coh})
\approx 2.5 \times 10^{-4} \ (\Delta v/10^{-3} \ v_{\rm
los})^{-1/2} \ a_{10}^{-1} \ r_{0.1}^{-1} \ R_{\rm c\, 13}
\end{equation}
and a cloud contribution to the mean gas density in the disk
of $<n_{\rm c}>\, \approx f_{\rm V,\, min} \, n_{\rm c}$. Since the radial
extent of the masing ring is typically $\gg \ell_{\rm coh}$,
one expects the clumped disk component to provide a nonnegligible
shielding column in the direction of the central continuum. For a cloud
located at a distance $\Delta r = r - \rin$ from the ring's
inner edge, the shielding column of intervening clumps can be approximated by
$N_{\rm cent} \approx \, <n_{\rm c}> \, \Delta r \approx (6
\vlos/\Delta v)^{1/2}(\Delta r/r) \, n_{\rm c} \, \Rc/a$. We can
eliminate $\Rc$ from this expression by fixing the effective maser
emission measure (eq. [16]) at $\xi \approx 1$, the minimum
value for efficient masing (EHM89). Then
\begin{equation}
R_{\rm c\, 13} = 5\, \xi\, \Delta v_5\, x_{-5}({\rm H_2O})\,
n_{\rm c\, 9}^{-2}\ ,
\end{equation}
and, for a ring around
a black hole of mass $M_{\rm bh}= 10^7\, M_7 \ M_{\odot}$,
\begin{equation}
N_{\rm cent} \approx 1.6 \times 10^{23} \, \xi \, M_7^{1/4} \,
r_{0.1}^{-1/4}\, \Delta v_5^{1/2}\, (\Delta r/0.5\, r)\, x_{-5}^{-1}({\rm
H_2O})\, n_{\rm c\, 9}^{-1} \, a_{10}^{-1}\ {\rm cm}^{-2}\ .
\end{equation}

We now combine the constraints given by the values of $N_{\rm
min}$, $N_{\rm max}$, and $N_{\rm cent}$. Using the parameters of
NGC 1068 as an example, we choose a radius near $r_{\rm out}$
($r = 1 \ {\rm pc}$) and plot these three column densities in an
$N_{\rm shield}$ vs. $x({\rm H_2O})$ diagram (Fig. 3). The
curves are labeled by the chosen cloud densities. For any
particular value of $\nc$, we focus on the portion of the
oblique ($N_{\rm cent}$) line that lies between the
corresponding horizontal ($N_{\rm min}$ and $N_{\rm max}$)
lines. This segment of the $N_{\rm cent}(\nc)$ curve corresponds to a
certain range of water abundances $x({\rm H_2O})$. Over that
range, dusty clouds located between $\rin \approx 0.65\ \pc$ and
$r=1 \ {\rm pc}$ can provide
sufficient shielding of the central continuum to render the
cloud at $r=1\ {\rm pc}$ molecular, but not so much
shielding that the cloud's temperature drops below $\sim 250
\ {\rm K}$. Since we chose a radius near $r_{\rm max}$, we
expect the cloud temperature to be close to $250 \ {\rm K}$: the
total shielding column must therefore be close to $N_{\rm max}$,
but if $N_{\rm cent}$ is lower than $N_{\rm max}$ one can attribute the
difference to the presence of an intercloud medium. Since the
requirement of a sufficiently high maser efficiency has already
been incorporated into the expression for $N_{\rm cent}$, the
indicated range would represent a self-consistent masing-ring
model if the required water abundance could be attained.

Figure 4 shows the water abundance and temperature as a
function of the distance $d$ from the irradiated surface of a
dusty slab that is placed at the adopted fiducial distance ($r =
1\ {\rm pc}$) from the center of NGC 1068.\footnote{Note that
dust-reprocessed radiation emitted within the slab is included
in the calculation of the gas thermal structure. The
contribution of this component is, however, expected to be
reduced in a realistic clumpy disk. The opposite limit, in which
the dust-reprocessed radiation is neglected altogether, was
adopted in the derivation of $N_{\rm max}$ in \S 3.} This
figure demonstrates that abundances as high as $x_{-5}({\rm H_2O}) \approx 1$
are attained in the irradiated clouds (consistent with
the values shown in Fig. 1 of Neufeld et al. 1994; note that Fig. 2
in that paper indicates significantly higher abundances, which
we, however, were unable to reproduce). In conjunction with
Figure 3, these results imply that essentially any cloud density
that lies in the optimal masing range could be accommodated by a
ring model that satisfies the indicated constraints. In
particular, the requirement $N_{\rm cent}[\nc,\, x({\rm
H_2O})] < N_{\rm max}(\nc)$ is evidently satisfied for
$x_{-5}({\rm H_2O}) \approx 1$ irrespective of the value of
$\nc$. (The insensitivity to the cloud density can be understood
from the fact that $N_{\rm max}$ and $N_{\rm cent}$ have a
similar dependence on $\nc$.) 

With all other parameters being
equal, lower-density clouds give rise to a higher shielding
column $N_{\rm cent}$ to the center. In a high-luminosity source like NGC 1068,
a relevant additional constraint is the {\em minimum} column
$N_{\rm rad}$ of dusty gas that would prevent the
ring from being dispersed by
the radiation-pressure force exerted by the central continuum. Since the
effective Eddington luminosity for dust opacity is $L_{\rm crit} \approx
10^{42} \ M_7 \ {\rm ergs} \ {\rm s}^{-1}$ (e.g., KK94),
and since the bulk of the continuum radiation is absorbed within a dusty 
column of $\sim 10^{21} \ {\rm cm}^{-2}$, we estimate
\begin{equation}
N_{\rm rad} \approx 10^{\rm 23} \ (L_{\rm bol}/10^{44} \, {\rm
ergs} \, {\rm s}^{-1})\ M_7^{-1} \ {\rm cm}^{-2}\ . 
\end{equation}
The value of $N_{\rm rad}$ for NGC 1068 is indicated in Figure 3 by the heavy
dashed line. If the typical cloud densities are $n_{\rm c\, 9}
\la 0.5$, then $N_{\rm max} \ga N_{\rm rad}$ and the gas in the maser
emission region can withstand ejection by the radiation pressure force.
We note from equations [6] and [19] that the condition $\Rc
< w_{\rm coh}$ implies, for $x_{-5}({\rm H_2O}) \la 1$ and $\xi
\ga 1$, that a cloud at $r=1\ {\rm pc}$ in this source must have
$n_{\rm c\, 9} > 0.07$. Clouds with $n_{\rm c\, 9} \approx 0.5$
would have $\Rc \approx 2 \times 10^{14}\ {\rm cm}$ and, even
with $a \approx 10$, would still satisfy $\ell_{\rm c} = a\, \Rc
\ll \ell_{\rm coh}$ ($\approx 2 \times 10^{17}\ {\rm cm}$). In
any case, we expect the disk to also contain nonmasing gas and
therefore to have an even greater inertia against
radiative acceleration. One indication of the presence
of nonmasing gas in masing AGN disks is provided by the
appearance of an absorption trough at the systemic velocity in
the maser spectra of sources like NGC 4258 (see \S 2.2). In the
current model, this absorption can arise in irradiated
gas beyond $r_{\rm out}$ whose temperature lies in the range
$\sim 100-250\ {\rm K}$. In the case of NGC 1068 it appears, making a rough
extrapolation of the curves in Figure 4, that an absorbing column
$\ga N_{\rm rad}$ could be maintained at
such temperatures even if the typical cloud
densities are such that the total column $N_{\rm max}(\nc)$ in
the masing region remains smaller than $N_{\rm rad}$. The
absorption optical depth is
maximized at $T \approx 200 \ {\rm K}$, and, for that
temperature and $\Delta v_5 \approx 1$ it is given by
\begin{equation}
\tau_{\rm absorb} \approx 5.9 \times 10^{-25} \ x_{-5}({\rm H_2O})\
N_{\rm absorb}\ ,
\end{equation}
where $N_{\rm absorb}$ is the absorbing hydrogen-nuclei column
(e.g., MM98). By setting $N_{\rm absorb} \ga
N_{\rm rad}$ ($\approx 2.4 \times 10^{23}\ {\rm cm}^{-2}$ for
NGC 1068) and adopting $x_{-5}({\rm H_2O}) \approx 1$, we
infer from equation (22) a systemic-velocity flux reduction
factor $\ga 1.2$, which seems to be consistent with the
data (see, e.g., Fig. 1 in Greenhill et al. 1996). Note,
however, that the extent to which the observer's line of sight to the center
intersects the $r \ga r_{\rm max}$ absorbing region in any given
source depends on the disk geometry and the viewing angle.

The properties of the masing ring in NGC 1068 that are implied
by this interpretation bear a remarkable resemblance to the inferred
characteristics of the CND in the Galactic center. In NGC 1068,
adopting $n_{\rm c\, 9} \approx 0.5$ as a typical cloud density,
we deduce $f_{\rm V,\, min} \approx 3 \times 10^{-4}$ (eq. [18]) and a
lower limit on the mass (considering only the masing gas between
$\rin$ and $r_{\rm out}$ and adopting a disk
half-thickness of $\sim 0.1\, r$) of $\sim 4 \times 10^3\
M_{\odot}$. Just like the NGC 1068 maser emission region, the
Galactic CND has the
appearance of a ring with a sharp inner edge (located at a
comparable distance from the center, $\rin \approx 1.5\ {\rm pc}$; the
outer radius of the CND has been inferred to be $\ga 7\ {\rm pc}$), it is
inclined with respect to the galactic plane (by $\sim
20-25^{\circ}$) and is slightly warped
(tilt angle varying in the range $\sim 10-25^{\circ}$; G\"usten
et al. 1987), and it exhibits a dominant circular motion (which in
the case of the CND corresponds to a nearly uniform rotational
speed of $\sim 110 \ \kms$) but with a nonnegligible velocity
dispersion. It also has a similar inferred mass ($\sim 10^4\
M_{\odot}$ within $4\ {\rm pc}$) and an indicated central black
hole of comparable mass (within a factor of $\sim 6$; see Eckart
\& Genzel 1997). It is particularly noteworthy that the CND 
appears to be highly clumped: HCN observations (e.g., Jackson et
al. 1993) have revealed that the bulk of the material is
characterized by hydrogen-nuclei
densities in the range $\sim 10^6-10^8\ \cmm$ and volume filling
factors $\sim 10^{-3}-10^{-2}$, corresponding to mean densities in
the range $\sim 10^4-10^5\ \cmm$, and by an area covering factor
$\sim 0.1$. In fact, there is no
observational evidence for the presence of a substantial
component of a low-density interclump medium in the CND. Also,
in both cases, there is evidence of mass inflow into the ring,
which on larger ($\sim 10^2-10^3\ {\rm pc}$) scales is associated with bar
streaming (e.g., Helfer \& Blitz 1995). In fact, as was first
noticed by Greenhill \& Gwinn (1997), the position angles of the
maser-ring axis and the nuclear CO bar in NGC 1068 differ by
only $\sim 20^{\circ}$. On the basis of these similarities one
can surmise that the Galactic CND would appear as a
water-maser ring akin to that in NGC 1068 if it had a comparably
bright high-energy central continuum.\footnote{Although one H$_2$O maser
was detected near the inner edge of the CND (Levine et al. 1995),
it appears to coincide with a luminous, reddened star and is
therefore likely to have a different origin than the NGC 1068 megamasers.} The
high-luminosity nonthermal emission
in NGC 1068, which is a signature of the currently active phase
of its nucleus, is evidently powered by accretion into the
center. As was pointed out by Helfer \& Blitz (1995), the main
difference in the circumnuclear molecular gas distribution
between NGC 1068 and the Milky Way is that the former appears to
have a large reservoir of gas (which could be the source of fuel
to the central engine) in the inner spiral arms, whereas in our own
galaxy there is a relative paucity of gas on similar scales.

The origin of the Galactic CND is still an open question, and if
similar structures indeed manifest themselves as extragalactic
megamaser disks, then its resolution could provide important
clues to the nature of accretion in AGNs (see \S 6). One of the
key issues is the presence of a well-defined inner edge at such
a comparatively
large distance from the center: interestingly, the presence
of such an edge has been indicated also in the H$_2$O maser
distributions of NGC 1068 and NGC 4258. If the inflowing gas
in the disk is broken up into distinct clumps as it approaches $\rin$ then 
the accretion efficiency could conceivably decrease
at lower radii, providing at least a partial
explanation for the presence of an ``edge'' at that
location. However, it appears that mass leaves the inner edge of
the CND in the form of ``streamers'' at about the same rate as
it is inferred to flow through the disk (Jackson et al. 1993).
It has been suggested that the presence of clumps and the
large measured velocity dispersion in the CND can be attributed
to tidal disruption (e.g., Genzel et al. 1994). The Roche critical density,
given (for an assumed molecular weight of 2.33) by
\begin{equation}
n_{\rm H,\, crit} \approx 3.5 \times 10^{11}\ M_7\ r_{0.1}^{-3}\ \cmm\ , 
\end{equation}
is $\sim 2 \times 10^9\ \cmm$ at the inner edge of the maser ring
in NGC 1068. If $\nc$ exceeds this value then the masing clouds in this object
could withstand tidal disruption and might therefore be relatively long lived.

The maser features that delineate the inner edge of the disk in
NGC 1068 likely amplify a background radio-continuum source,
as has been inferred to be the case for the systemic masers in
NGC 4258 (see \S 2.2). In fact, Gallimore et al. (1997) obtained
radio images of a parsec-scale disk in NGC 1068 that they
interpreted in terms of a hot, ionized gas at the inner edge of the nuclear
disk. It is therefore possible that the ring inner edge itself
is the source of the background radiation. It is interesting to
note in this connection that the Galactic CND also exhibits an
ionized, radio-emitting inner rim (e.g., Genzel et
al. 1994). In fact, only part of the CND's inner circumference (notably,
the so-called Western Arc) is manifested in this fashion, which
has been attributed to absorption of the central ionizing radiation
in certain directions by infalling matter within the central
cavity (Jackson et al. 1993). A similar explanation may perhaps
be relevant to the fact that only one quadrant of the inner rim
of the NGC 1068 ring has a measurable maser flux. The
confinement of the maser emission to the vicinity of the disk
inner edge could reflect the fact that the background radiation is
intercepted and absorbed within a narrow radial range near $\rin$.
It is nevertheless puzzling why one does not observe enhanced
maser emission near the centerline. In the case of a planar disk
and a $90^{\circ}$ viewing angle, both background- and
self-amplification by clouds are expected to contribute to the
flux near the systemic velocity on account of the comparatively
large velocity coherence length near the centerline. (As
noted above, the emission originating right along and in the
immediate vicinity of the centerline
would be subject to absorption by the nonmasing gas at $r > r_{\rm out}$.)
The absence of a flux peak near the systemic velocity may be due
to the disk's departure from planarity or velocity coherence, or
to the viewing angle being sufficiently smaller than
$90^{\circ}$. It is, however, also conceivable that
the total shielding column through the disk is
large enough for all the ionizing continuum radiation to be
absorbed near $\rin$ (see MM98). In that case a
continuum-irradiation excitation of the high-velocity features would require
raising the masing gas above the midplane as originally
envisioned by Neufeld \& Maloney (1995).

Before concluding our discussion of this scenario it is of
interest to inquire whether the ``minimalist'' model that we
considered for NGC 1068 could also be applied to NGC 4258. Figures 5
and 6 are the analogs of Figures 3 and 4, respectively, and are
based on the same assumptions, except that in this case we
use the inferred spectrum of NGC 4258 and set $r = 0.25 \ {\rm
pc}$ (which is close to $r_{\rm out}$ in this source). We do not
plot the value of $N_{\rm rad}$ (eq. [21]) since
radiation-pressure acceleration is not an issue in this
low-luminosity ($L_{\rm bol} \approx 10^{42}\ {\rm ergs \
s^{-1}}$) object. We find that, as before, the photoionization-induced
water abundance does not rise above $x_{-5}({\rm H_2O}) \approx 1$,
and that, for this value, $N_{\rm cent}(\nc)$ is significantly
larger than $N_{\rm max}(\nc)$. In other words, for the minimum
cloud space density (eq. [17]) required to explain the
high-velocity features in terms of overlapping maser clumps, the implied
cloud self-shielding of the central continuum is too high to
allow the masing region to extend as far as it is observed to do. 

A possible resolution of the above difficulty is that $x({\rm
H_2O})$ is, in fact, greater than $\sim 10^{-5}$. One potential
source of water in dusty clouds is
the evaporation of icy mantles of grains that become exposed to the
central radiation field and are heated to a temperature $\ga
100 \ {\rm K}$. Ice could form on the surfaces of grains that
are embedded in a cold, molecular disk through the hydrogenation of oxygen
(Brown, Charnley, \& Millar 1988). Infrared observations of star-forming
regions in dense molecular clouds have revealed that $\sim 10\%$ of the
elemental oxygen is in water ice: in that case $x({\rm H_2O})$
could be $\sim 6 \times 10^{-5}$ (Ceccarelli,
Hollenbach, \& Tielens 1996). Brown et al. (1988) originally
hypothesized that this mechanism could yield water abundances as
high as $\sim 5 \times 10^{-4}$. For $x_{-5}({\rm H_2O}) \approx
6$, Figure 5 indicates that a planar disk model could be
self-consistent only for $n_{\rm c\, 9} > 1$. For cloud
densities of that order one can infer from Figure 6 that the column of
absorbing gas at $r > r_{\rm max}$ would be $\la 10^{24}\ {\rm
cm}^{-2}$, which, by equation (22) with $x_{-5}({\rm H_2O}) \approx
6$, would yield a flux reduction factor $\sim 30$. This is
consistent with the systemic-velocity value of $\sim 10$
measured in this source. For $\xi \approx 1$, $x_{-5}({\rm H_2O})
\approx 6$, and $n_{\rm c\, 9} \ga 1$, the characteristic cloud radius
would be $r_{\rm c\, 13} \la 1$ (eq. [19]). We conclude that,
if ice evaporation as invoked in collapsing protostellar
envelope calculations is also a viable mechanism in molecular
accretion flows in AGNs, then the ``minimalist'' planar ring
model could in principle also apply to a source like NGC
4258. If the mantle-evaporation mechanism is for some reason
inapplicable then NGC 4258 might still represent a clumpy ring,
but in order for the maser emission to be excited by continuum irradiation
the maser clouds could not all lie in the same plane.

\section{Clouds Uplifted by a Disk-Driven Wind}

In \S 4 we discussed irradiated clumpy rings and showed that, while they may
adequately explain the maser distributions in objects like NGC
1068 and perhaps even NGC 4258, certain aspects of the observations may
be easier to interpret if the masing clouds are not strictly
coplanar. Since the maser disks in both of these sources in
fact appear to be slightly warped (a feature that they share
also with the Galactic circumnuclear ring), one is led to
examine plausible physical mechanisms that could give rise to
a non-coplanar distribution of masing clouds.
As we noted in \S 1, previously proposed warping
scenarios were constructed for homogeneous disk models. Although
these mechanisms would presumably remain applicable if the disk
contained a turbulent velocity field, it is less clear how
efficiently they would operate in a highly inhomogeneous disk
with a distinct, low--filling-factor cloud component. In this
section we address the issue of a non-coplanar cloud distribution
and propose an alternative to the warped-disk interpretation of
continuum-irradiated megamaser sources. We suggest that the
maser features represent diamagnetic molecular clouds that are
uplifted from the surface of a circumnuclear accretion disk by a
dusty hydromagnetic wind. In this picture the wind is driven
centrifugally along rotating magnetic field lines that are
anchored in the disk (Blandford \& Payne 1982). Dust from the
outer, molecular regions of the disk is carried out with the gas
and survives destruction by the intense central continuum
along wind streamlines whose footpoints in the disk lie beyond
the dust sublimation radius $r_{\rm sub}$. KK94 showed
that such a wind could produce the obscuration
of the central continuum source that has been attributed to the 
phenomenological molecular ``torus'' that is inferred to
surround the broad emission-line region of Seyfert 2 galaxies
(e.g., Antonucci 1993). The incorporation of clouds into such a
wind has previously been considered in the context of models of
BELR clouds in Seyfert galaxies and QSOs (Emmering, Blandford,
\& Shlosman 1992;
Bottorff et al. 1997), EUV and X-ray absorption-line clouds in BL
Lacertae objects (K\"onigl et al. 1995; Kartje et al. 1997), and
broad absorption-line QSOs (de Kool \& Begelman 1995; Kartje \&
K\"onigl 1997, 1998). Although in this paper we do not specifically address
the origin of the clouds, the internal densities and space
densities of the masing clumps that we model are consistent with
the parameters inferred for BELR clouds (e.g., Baldwin et
al. 1995; Arav et al. 1997).

In applying the uplifted-cloud scenario to H$_2$O megamaser
sources we show that the necessary shielding
from the high-energy central-continuum photons can be provided for
comparatively massive clouds that initially move on nearly
circular trajectories close to the disk surface. Since the
velocity field of these clouds is well approximated by the rotation law of the
underlying disk, it follows from the considerations of \S 2 that this
scenario can account for the detection of H$_2$O 
maser features in Seyfert galaxies that are viewed at
large angles to their symmetry axes. In this picture the
apparent warp in the maser
distribution may result from an observational selection effect: a cloud
uplifted at a given radius would attain optimal masing
conditions at a height large enough that shielding by the dust in the
intervening wind and clouds is not too high, but not so large
that the shielding column or the cloud density become too low
for efficient masing. Both the shielding column to the center
and the cloud density (which is determined by the confining ambient pressure)
drop rapidly with height in centrifugally driven outflows of the
type that we envision. Therefore, for typical wind mass-outflow
rates, the maser emission region would be confined to the
immediate vicinity of the disk surface. Since the
maser-shielding column in this interpretation is directly related to the
equivalent X-ray--absorbing hydrogen column density, both being
associated with the disk-driven wind (see KK94), this model also readily
accounts for the inferred correlation between the
probability of detecting H$_2$O maser emission in a Seyfert
galaxy and the magnitude of its X-ray--absorbing column (Braatz
et al. 1997).

\subsection{Cloud Dynamics}

In our proposed picture, dense clouds fragmented from the disk are
uplifted by the wind ram pressure and transported along the
streamlines (with radiation pressure possibly contributing to the acceleration
once the clouds get exposed to the central continuum source; see
Kartje \& K\"onigl 1997, 1998). A nominal estimate of the
maximum cloud mass that can be ejected from the disk at a radius $r$ may be
obtained by balancing the wind ram-pressure and tidal
gravitational forces (see eq. 25]). This gives
\begin{equation}
M_{\rm c,\, max} \approx \pi \dot
M_{\rm w} R_{\rm c}^2/
(GM_{\rm bh} r)^{1/2} = 9.8 \times 10^{23}\ \dot M_{\rm
w,-3} \ R_{\rm c\, 13}^2 \ 
M_7^{-1/2} \ r_{0.1}^{-1/2} \ \ {\rm g} \, ,
\end{equation}
where $\dot M_{\rm w} = 10^{-3}\, 
\dot M_{\rm w,-3} \ M_\odot \ {\rm yr}^{-1}$ is the mass outflow
rate in the wind. The clouds can be assumed to be pressure-confined by the wind
magnetic field. The approximate cloud
trajectories under the joint action of the wind--cloud drag
force and the gravitational field of the central mass can be
obtained by solving the equation
\begin{equation}
\frac{d{\bf v_{\rm c}}}{dt} \approx -\frac{GM}{r^2}{\bf \hat{r}} + 
\frac{\pi R_{\rm c}^2 \rho_{\rm w}}{M_{\rm c}} \left | {\bf v_{\rm w}} - 
{\bf v_{\rm c}} \right | ({\bf v_{\rm w}} - {\bf v_{\rm c}}) \, ,
\end{equation}
where ${\bf v}_{\rm c}, \ M_{\rm c}, \ {\bf v}_{\rm w}, \ {\rm and} \ 
\rho_{\rm w}$
are the cloud velocity, cloud mass, wind velocity, and wind mass density,
respectively.  We assume that the cloud density is determined by the local
wind magnetic field strength $B_{\rm w}$ ($n_{\rm c} \propto B_{\rm
w}^2$ for $T$ nearly constant).  We further assume that the mass
of each cloud does not change and that the clouds maintain a roughly
spherical shape (i.e., $R_{\rm c} \propto B_{\rm w}^{-2/3}$).

In view of the minimum shielding required for keeping the clouds
molecular (see Fig. 1), the masing clumps must be located at $r
> r_{\rm sub}$ behind a sufficiently large column of dusty wind
and uplifted clouds. The precise geometry and kinematic
properties of the maser emission region depend both on
the cloud characteristics (sizes, masses, and ejection radii)
and on the dynamical properties of the wind. In particular, the
wind mass outflow rate $\dot M_{\rm w}$ is a key parameter. In
sources with a high value of $\dot M_{\rm w}$ the
dust shielding might be effective up to comparatively high latitudes and
the projected locus of efficient maser emission within the wind
would take the form of a narrow strip (whose width increases
with height above the
disk) that extends along the innermost dusty streamline (see
Fig. 7$a$). In this case the line-of-sight velocities of high-latitude maser
features would be expected to drop with distance from the rotation
axis in a sub-Keplerian fashion even if the equatorial disk were
Keplerian (see Fig. 7$b$). By contrast, in winds with a lower mass
outflow rate, which are less effective at shielding the central
continuum, the projected locus of efficient masing would be a narrow
strip extending along (and right on top of) the disk surface (Fig. 7$a$).
To insure that their trajectories lie within the most highly
shielded wind region close to the disk surface, the masing
clouds in this case would need to be fairly heavy ($M_{\rm c}
> M_{\rm c,\, max}$; see eq. [24]). The motions of such
relatively massive clouds would closely mirror the ambient velocity
field near the disk surface, and hence, in the case of a Keplerian disk, their
line-of-sight velocities would satisfy $v_{\rm los} \propto
r^{-1/2}$ (Fig. 7$b$). The latter alternative is evidently the
one that best describes the basic
morphology and kinematics of the H$_2$O masers in
a source like NGC 4258. 
In light of the arguments presented in \S 2,
wind-uplifted clouds that overlap along the line of sight could
in principle account for the observed flux level and variability
of the maser features in this source.

\subsection{A Clumpy Disk-Driven Wind Model}

To flesh out the above scenario we now incorporate the
dynamical constraints on the uplifted maser clouds into a
``generic'' model of a disk-driven hydromagnetic wind. As we
discussed in \S 5.1, the maser observations in an object like NGC
4258 are best described in the context of this
picture in terms of clouds of masses $M_{\rm c} = (4\pi/3)\, (1.4 \,
m_{\rm H})\, \Rc^3\, \nc \ga 10\, M_{\rm c,\, max}$ (where
the factor 1.4 accounts for the assumed composition of molecular
hydrogen with a 20\% helium abundance). We thus
parametrize $M_{\rm c} = \eta \, M_{\rm
c,\, max}$, with $\eta = 10 \, \eta_{10}$. Another relation
between the cloud and wind properties is provided by the
pressure confinement condition. We assume that the clouds are
diamagnetic (i.e., that they are not penetrated by the ambient
magnetic field) and impose a balance between their internal (mostly thermal)
pressure and the wind (mostly magnetic) pressure. This can be
written as
\begin{equation}
\frac{B_{\rm w}^2}{8\, \pi} \approx 0.6\, \nc\, k\, T_{\rm c}\ 
\end{equation}
(where the factor 0.6 accounts for the assumed composition).

To make the model self-contained, we assume that the wind
extracts most of the angular momentum of the accreted
matter. This allows us to relate the magnetic field strength in
the wind to the mass accretion rate $\dot M_{\rm acc}$ through
the disk. Using the angular momentum conservation relation for a
thin, equatorial disk, $\dot M_{\rm acc}\, v_{\rm K}/8\, \pi\,
r^2 \approx B_{\rm z}\, |B_{\phi,\, {\rm s}}|/4\, \pi$ (where
$B_{\rm z}$ and $B_{\phi,\, {\rm s}}$ are, respectively, the vertical component
and the azimuthal surface component of the disk magnetic field),
we obtain
\begin{equation}
\frac{B_{\rm w}^2}{8\, \pi} = \frac{\dot M_{\rm acc}\, v_{\rm
K}}{16\, \pi\, \beta\, r^2}\ ,
\end{equation}
where the parameter $\beta = B_{\rm z}\, |B_{\phi,\, {\rm s}}|/B_{\rm w}^2$
depends on the detailed magnetic field configuration near the
disk surface. In the cold, radially self-similar wind model of Blandford
\& Payne (1982), which has been employed in several of the
applications cited at the beginning of this section, one can express
$\beta$ in terms of the self-similarity variables $\kappa$ (the
normalized mass-to-flux ratio), $\lambda$ (the normalized total
specific angular momentum), and $\xi_0^{\prime}$ (the
radial-to-vertical field-component ratio at the disk surface) as $\beta =
\kappa \, (\lambda - 1) / [\kappa^2\, (\lambda - 1)^2 + (1 +
{\xi_0^{\prime}}^2)]$. This yields $\beta \approx 0.32$ for the
wind solution used in Figure 7 (see Table 1 in KK94). We thus
parametrize $\beta = 0.3\, \beta_{0.3}$.

To complete the specification of the model we need to relate
the wind mass outflow rate (which appears in eq. [24] for
$M_{\rm c,\, max}$) to the disk mass accretion rate
(which appears in eq. [27]). We parametrize $\dot M_{\rm w} =
\delta\, \dot M_{\rm acc}$ and choose $\delta_{0.1} = \delta/10$ as
a representative normalization factor. Using the adopted fiducial values
of $\beta$, $\eta$, and $\delta$, and parametrizing the
temperature in the masing clouds by $T_{\rm c\, 500} = (T_{\rm c} /
500\ {\rm K})$, we can combine the above expressions to get
\begin{equation}
\Rc \approx \frac{36\, \pi\, \beta\, \eta\, \delta\, k\,
T_{\rm c}\, r^2}{7\, G\, M_{\rm bh}\, m_{\rm H}} \approx 1.4
\times 10^{13} \,
\beta_{0.3}\eta_{10}\delta_{0.1}T_{\rm c\, 500}r_{0.1}^2 M_7^{-1}\
{\rm cm} \ .
\end{equation}
It is remarkable that the typical cloud radius obtained in this
fashion for a source like NGC 4258 agrees with the value
inferred from the requirements of
efficient maser pumping (see eq. [19]). If we explicitly impose the
condition that the maser efficiency factor $\xi$ (eq. [16]) be $\ga
1$, we can directly estimate the characteristic mass accretion rate through the
disk,
\begin{equation}
\dot M_{\rm acc} \approx 2.7 \times 10^{-2}\, 
\beta_{0.3}^{1/2}\eta_{10}^{-1/2}\delta_{0.1}^{-1/2}\xi^{1/2}\Delta
v_5^{1/2}x_{-5}({\rm H_2O})^{-1/2}T_{\rm c\, 500}^{1/2}r_{0.1}^{3/2}\
M_{\odot} \ {\rm yr}^{-1}\ .
\end{equation}
Interestingly, this value is compatible with the ADAF
interpretation of the NGC 4258 spectrum (see \S 1).\footnote{As
we show in the Appendix, if a centrifugally driven wind carries
away much of the disk angular momentum, then the ADAF
interpretation can also be compatible with the radiation-induced warping
instability model for this source.} In
conjunction with equation (27) we then deduce
\begin{equation}
B_{\rm w} \approx 4.4 \times 10^{-2}\, 
\beta_{0.3}^{-1/4}\eta_{10}^{-1/4}\delta_{0.1}^{-1/4}\xi^{1/4}\Delta
v_5^{1/4}x_{-5}({\rm H_2O})^{-1/4}T_{\rm c\, 500}^{1/4}M_7^{1/4}r_{0.1}^{-1/2}\
{\rm G}\ .
\end{equation}
The toroidal field component $|B_{\phi,\, {\rm s}}|$ would be smaller
than this value: using again the Blandford \& Payne (1982)
similarity variables, we deduce $|B_{\phi,\, {\rm s}}|/B_{\rm w} =
\kappa \, (\lambda - 1) / [\kappa^2\, (\lambda - 1)^2 + (1 +
{\xi_0^{\prime}}^2)]^{1/2}$, which for the wind solution used in
Figure 7 is $\sim 0.55$. Hence, at a distance $r_{0.1} \approx
2$ in NGC 4258 (where $M_7 \approx 3.5$), a line-of-sight field component of
$\sim 2 \times 10^{-2}\ {\rm G}$ is predicted {\rm in the
wind}. This is lower than the upper limit of $0.3 \ {\rm G}$
inferred from circular polarization measurements in the maser
feature at that location (Herrnstein et al. 1998c), although we note
that, under our assumption
of diamagnetic clouds, the magnetic field amplitude in the
masing clumps might in fact be lower than the
estimate (eq. [30]) for the homogeneous outflow component.

As a self-consistency check, we evaluate the shielding
column associated with the dusty wind itself.
Assuming that the wind is atomic and has a vertical
speed $v_{\rm w,\, z} = \chi \, C(T_{\rm w})$, where $\chi = 3
\, \chi_3$ and $C(T_{\rm w})$
is the isothermal speed of sound corresponding to the wind
temperature $T_{\rm w} \approx 10^4\ {\rm K}$,
and approximating $\dot M_{\rm w}
\approx 2\, \pi\, r\, (1.4\, m_{\rm H})\, v_{\rm w,\, z}\,
N_{\rm H,\, w}$, we infer a shielding column
\begin{equation}
N_{\rm H,\, w} \approx 1.5 \times 10^{22}\, 
\beta_{0.3}^{1/2}\eta_{10}^{-1/2}\delta_{0.1}^{1/2}
\chi_3^{-1}\xi^{1/2}\Delta
v_5^{1/2}x_{-5}({\rm H_2O})^{-1/2}r_{0.1}^{1/2}(T_{\rm w}/20\,
T_{\rm c})^{-1/2}\ {\rm cm}^{-2}\ .
\end{equation}
For $r \la r_{\rm out}$ this value should not exceed the
maximum shielding column $N_{\rm max}$ for collisional pumping (see \S 3).
By substituting the observed value of $r_{\rm out}$ in
NGC 4258 and using the fiducial values of all the other
parameters in the expression (31), we obtain $N_{\rm H,\, w} \approx 2.5 \times
10^{22}\ {\rm cm}^{-2}$, which is approximately equal to
the value of $N_{\rm max}$ shown in Figure 5 for the corresponding
cloud density ($n_{\rm c\, 9} \approx 2$), assuming $x_{-5}({\rm H_2O}) \approx
1$ (see Fig. 6). This suggests that in this case the interclump
medium (i.e., the dusty wind) may provide most of the requisite
shielding of the masing gas from the incident high-energy
radiation, whereas the clumped component (i.e., the uplifted
clouds) mainly contributes the high-density gas that is needed
for the collisional pumping of the masers.
All in all, even though a comprehensive dynamical
calculation of a clumpy wind is evidently required to fully
establish the applicability of this model, the results of
our simple analysis indicate that this is likely to be a viable explanation
of the maser distribution in a source like NGC 4258.

\section{Summary and Discussion}

Water megamaser emission in AGNs is typically detected in Seyfert 2-like
galaxies that are viewed at a large angle to the symmetry axis. Recent VLBI
imaging observations of several of these sources reveal that the masers
generally appear as discrete features that lie in a thin disk of radius $r \sim
0.1-1 \ {\rm pc}$ and radial extent $\Delta r \la r$. Spectroscopic data
indicate that the masers rotate with a nearly Keplerian velocity distribution,
with high-velocity red- and blue-shifted satellite features appearing on
either side of the projected disk diameter (the midline). The masers are
believed to be pumped collisionally.

By comparing the maser-emitting volume that produces the satellite features
in a source like NGC 1068 with the maximum volume permitted by
velocity-coherence considerations, and on the basis of the spectral and
temporal behavior of the maser flux in a source like NGC 4258, we argue that,
independent of the exact maser excitation mechanism, the emitting gas is likely
clumped. The clumps could correspond to velocity irregularities in a turbulent
medium, but in our specific models we assume that the maser features in fact
represent distinct dense clouds. We show that even two overlapping filamentary
clumps near the midline of a Keplerian disk that are separated from each other
(along the line of sight) by less than the velocity-coherence length can
account, through self amplification, for the entire flux of a maser feature in
sources like NGC 1068 and NGC 4258. The observed flux variability can then be
attributed to relative motions that bring the clumps into or out of alignment
with each other along the line of sight. The clump overlap probability is
maximized for edge-on viewing of the disk, especially in the case of saturated
emission (for which the minimum cloud separation that is required to account
for the observed high-velocity maser fluxes can approach the coherence length):
we point out that this could explain the strong observational bias in favor of
detecting H$_2$O maser emission in galaxies that are observed at a large angle
to the disk rotation axis.

The interpretation of H$_2$O maser features in terms of aligned dense clumps
has previously been proposed for Galactic sources (Deguchi \& Watson 1989;
EMH91). One example is provided by W49N, the most luminous Galactic water-maser
source, whose brightest features have fluxes of a few times $10^4$ Jy (e.g.,
Gwinn 1994). Interestingly, these fluxes are within about an order of magnitude
of the peak fluxes of the high-velocity maser features in sources like NGC 4258
and NGC 1068 when the distance is scaled from $\sim 11 \ {\rm kpc}$ to $\sim 6
\ {\rm Mpc}$ and $\sim 15 \ {\rm Mpc}$, respectively. Furthermore, the bright
outbursts in W49N have been commonly attributed to chance alignments of
individual maser features (e.g., EMH91), and it has even been argued that they
involve distinct clouds (Boboltz et al. 1998).  All in all, NGC 4258 resembles
maser outbursts in W49N both in terms of the flux level and variability. 
Another example is provided by the H$_2$O maser source in Orion, where
polarization measurements of a strong flare have been interpreted in terms of
aligned clumps in a rotating disk that is viewed edge on (e.g., Matveenko,
Graham, \& Diamond 1988; Abraham \& Vilas Boas 1994). A similar disk
interpretation has also been proposed for other water-maser sources associated
with young stellar objects (e.g., Fiebig et al. 1996; Torrelles
et al. 1998). These examples suggest that the basic properties of extragalactic
water-megamaser sources may be quite similar to those of their bright Galactic
counterparts even if the circumstances under which they arise are unique to the
AGN environment.

In this paper we focus attention on a class of models that attribute the maser
emission to heating by the central continuum source. In order for water
molecules to survive the intense irradiation at the observed positions of the
masers, shielding by a dusty medium of column density $N_{\rm shield} \ga
10^{22}-10^{23} \ {\rm cm}^{-2}$ is required. Gas densities in the range $\sim
10^8 - 10^{10} \ {\rm cm}^{-3}$, a temperature $\ga 250 \ {\rm K}$, and a
fractional water abundance $\sim 10^ {-5} - 10^{-4}$ are also needed for
optimal masing. In our interpretation of the megamaser disks we identify the
outer radius $r_{\rm out}$ of the maser emission region with the shielding
column $N_{\rm max}$ above which the temperature drops below $\sim 250\ {\rm
K}$ and collisional pumping of water molecules ceases. We incorporate the above
constraints into a ``minimalist'' scenario for AGN maser disks, in which the
disk represents a low-mass ring that consists of a dominant cloud component and
possibly also of a more tenuous intercloud medium. In the simplest
representation of this model the radiation reaches the clouds through the plane
of the ring: we find that such a flat-disk model is consistent with the
observations in NGC 1068 and may also be applicable to NGC 4258 if the water
abundance exceeds (due, e.g., to the evaporation of icy grain mantles) the
value determined by equilibrium photoionization-driven chemistry. The required
gas densities and space densities of the clumps in this scenario are consistent
with the values inferred for BELR-type clouds in AGNs.

We note the striking similarity between the ring properties implied by this
interpretation of the tilted maser ring in NGC 1068 and those attributed to the
circumnuclear disk (CND) in the Galactic center. Various suggestions have been
considered for the origin of the CND and of its well-defined inner edge (e.g.,
Jackson et al. 1993), and the study of analogous structures in AGN megamaser
disks might help resolve these issues. In the case of the CND there is evidence
for ongoing accretion through the disk at a rate of $\sim 3 \times 10^{-2}\
M_{\odot}\ {\rm yr}^{-1}$, which is similar to the values inferred in some of
the megamaser disks. Furthermore, in the Galactic case it has been argued on
the basis of several dynamical considerations that the ring is a short-lived
($< 10^5\ {\rm yr}$) structure (e.g., Genzel et al. 1994). If a similar
conclusion applies to the extragalactic magamaser rings then this could have
important implications to our picture of how mass accumulates in the centers of
AGNs. In particular, could it be that, on the scales of the maser emission
regions, most of the mass flows to the center through short-lived,
comparatively low-mass disks whose axes may be significantly tilted with
respect to the local symmetry axis (as determined, for example, by the nuclear
jet)? If most of the mass in such disks is in the clumped component, as
inferred in the CND and consistent with the above interpretation of the
megamaser disk in NGC 1068, how exactly is the angular momentum of the
inflowing gas transported away? In a source like NGC 1068, does there also
exist a massive circumnuclear disk that is roughly perpendicular to the inner
radio-jet axis, as has been inferred from various other observations (e.g.,
Antonucci 1993; Gallimore et al. 1996b, 1997; Tacconi et al. 1997)? If a
distinct massive disk indeed exists in this archetypal object, does it interact
dynamically with the maser ring? Could the simultaneous presence of edge-on
disks and tilted rings in Seyfert 2-like galaxies be relevant to the puzzling
finding of reprocessed K$\alpha$ lines in Seyfert 2 and narrow emission-line
galaxies (Turner et al. 1998)? Should the ring picture be found to adequately
describe the maser observations, then these and other related questions would
need to be critically addressed.

We also consider the case in which the disk is too massive to permit the
central continuum radiation to reach the masing clumps through its plane. The
apparent warp seen in the maser distributions of both NGC 4258 and NGC 1068
provides observational support to the notion that the masing gas intercepts the
incident radiation above the midplane. We present, however, a scenario that,
unlike the homogeneous-disk model of Neufeld \& Maloney (1995), does not
require the disk to be physically warped. Our proposed model is based on the
hydromagnetic disk-driven wind scenario we had previously applied to the
interpretation of the molecular ``tori'' in Seyfert 2-like galaxies and to
other phenomena in AGNs (e.g., KK94, Kartje \& K\"onigl 1997). In this picture,
the wind uplifts (by its ram pressure) and confines (by its magnetic pressure)
dense, diamagnetic clouds that form near the disk surface. The maser emission
originates in the dense clouds once they become exposed to the central
continuum radiation, with the dusty regions of the wind providing much of the
requisite shielding. The apparent warp may then represent an observational
selection effect induced by the strong vertical density stratification that
characterizes centrifugally driven winds. We analyze the ingredients of a
self-contained ``generic'' model of a clumpy wind that transports the bulk of
the angular momentum of the underlying accretion disk. Using representative
parameters we show that this model can yield optimal masing conditions in a
source like NGC 4258. We also demonstrate that the mass accretion rate implied
by a hydromagnetic wind-driving disk model of this type is consistent with the
ADAF interpretation of the spectrum of this source (Lasota et al. 1996).

The spiral-shock model of Maoz \& McKee (1998) provides an alternative to the
continuum-irradiation picture of megamaser disks. The shock scenario does not
conflict with the discrete-clumps interpretation presented in this paper since
the maser emission in the clumps could in principle be excited by shocks. In
fact, the relative motion of distinct clumps is still the most natural
explanation of the strong flux variability exhibited by a source like NGC 4258.
One attractive feature of the spiral-shock model is that it readily accounts
for the apparent tendency of the redshifted high-velocity maser features in the
existing sample of megamaser sources to dominate over the blueshifted ones.
There are, however, reported cases where the blueshifted features dominate
(e.g., NGC 3079; Trotter et al. 1998), including a case (NGC 5793) where the
flux of a redshifted feature decreased from being twice as large as the
blueshifted maser flux to below the detection threshold on a time scale of
$\sim 1$ month (Hagiwara et al. 1997). In these instances it is likely that the
effects of cloud self-amplification, rather than the geometry of any shocks
that might be present, control the appearance of the source. This conclusion is
supported, in the case of NGC 3079, by the apparent fading and lighting-up of
certain maser features between consecutive observations, and it remains valid
even if the maser emission in this source is excited by some other mechanical
means (e.g., by the impact of a misaligned jet on the disk, as proposed by
Trotter et al. 1998) rather than by spiral shocks. Further insights into the
excitation mechanism of the H$_2$O megamasers might be provided by studies of
other masing molecules, such as OH. In fact, it is now becoming clear that at
least some of the OH maser emission detected in extragalactic sources is
associated with the nuclei of active galaxies (e.g., Trotter et al. 1997;
Lonsdale et al. 1998), and in some cases (e.g., NGC 1068; Gallimore et al.
1996a) it even appears to originate in the same material that produces the
water maser radiation. These studies may be expected to lead to a comprehensive
model of the megamaser emission in AGNs and could shed new light on the nature
of the associated disks.

\acknowledgments

We thank G. Ferland for assistance with the use of {\em CLOUDY},
R. Antonucci, R. Barvainis, G. Ciolek, C. Gwinn, D. Hollenbach,
J. Moran, and W. Watson for valuable conversations and
correspondence, and the anonymous referee for insightful
comments. A.K. also thanks T. Mazeh and H.
Netzer for helpful discussions and hospitality at Tel Aviv University during
part of this work. This research was supported in part by NASA grants NAG
5-2766 \& NAG 5-3687 (U.C.) and NAG 5-3010 \& NAG 5-7031 (U.Ky.).

\newcommand{\bc}{\begin{center}}
\newcommand{\ec}{\end{center}}
\setcounter{equation}{0}
%\vspace{-3ex}
\renewcommand{\theequation}{A{\arabic{equation}}}
\bc {\bf APPENDIX} \ec
%\vspace{-6ex}
\bc {\bf RADIATION-INDUCED WARPING INSTABILITY IN A
HYDROMAGNETIC WIND-DRIVING ACCRETION DISK} \ec
%\vspace{-3ex}

In \S 5 we discuss a clumpy, accretion disk-driven hydromagnetic
wind model for water megamaser sources. We show that,
under the assumption that a significant fraction of the
disk angular momentum is carried away by the wind, the implied mass
accretion rate is consistent with that inferred from the ADAF
interpretation of the spectrum in NGC 4258 (Lasota et
al. 1996). In contrast, the irradiated viscous-disk scenario considered by
Neufeld \& Maloney (1995) implies an accretion rate that is
significantly smaller than the ADAF value. In this Appendix we
point out that a similar wind-driving accretion-disk model can
also resolve the discrepancy between the ADAF interpretation
and the radiation-induced warping-instability scenario for the
origin of the apparent warp in this source.\footnote{It is
worth noting that the apparent warps that have so far been observed in
megamaser disks (as well as in the Galactic CND) are
comparatively mild and do not directly support the proposition
that the radiation-induced warping instability could account for
the collimated ``ionization cones'' found in Seyfert 2 galaxies
(see Begelman \& Bland-Hawthorn 1997) .} One should,
however, bear in mind that the warping-instability model for homogeneous
disks is clearly distinct from the uplifted-cloud interpretation
of the apparent warp outlined in \S 5.

In the radiative-instability mechanism, first studied by Petterson
(1977) and more recently elaborated on by Pringle (1996) and
Maloney et al. (1996b), a disk illuminated by a central radiation
source becomes unstable to warping if it is optically thick to
both absorption and emission. The origin of the warp is the
torque exerted locally on the disk by the reemitted radiation, which is
directed normal to the disk surface (in contrast to the absorbed
flux, which, being radial, produces no torque).\footnote{An analogous
mechanism, involving the torque exerted by a wind evaporated
from the irradiated disk surface, was proposed by Schandl \&
Meyer (1994).} The net torque
on any given annulus of the disk is nonzero if the annulus is
warped, and this would generally occur if there is a radial
gradient in either the tilt angle or the angle of the line of
nodes of the disk. Pringle (1996), using a local stability
analysis, showed that the instability in a steadily accreting source
is expected to arise at radii greater than
\begin{equation}
r_{\rm warp, \, min} = \zeta \ (\psi/\epsilon)^2 \ R_{\rm s} \, ,
\end{equation}
where $\psi=\nu_2/\nu_1$ is the ratio of the effective $(r,z)$
and $(r,\phi)$ viscosities, $\epsilon = L_{\rm bol} / \dot M_{\rm
acc} \, c^2$ is the radiative efficiency, $R_{\rm s} = 2 \, G \,
M/c^2$ is the Schwarzschild radius, and the constant $\zeta$ is
$\sim 8 \, \pi^2$. Maloney et al. (1996b) applied the radiation-induced warping
instability to the interpretation of the apparent warp in NGC
4258 (e.g., Herrnstein et al. 1996). They noted,
however, that this interpretation is inconsistent with the 
advection-dominated accretion model for this source. The origin
of the difficulty can be seen by
estimating the radiative efficiency in equation (A1) on the
assumption of advection-dominated accretion. Taking
$L_{\rm bol} \approx 10^{42} \ {\rm ergs \ s^{-1}}$ and adopting a viscosity
parameter $\ga 0.1$, one gets $\epsilon \la 2 \times
10^{-2}$. Using this estimate and the fiducial value of $\zeta$
yields $r_{\rm warp, \, min} \ga 2 \times 10^5 \ 
\psi^2 \ R_{\rm s}$, which exceeds $r_{\rm min} \approx 4
\times 10^4 \ R_{\rm s}$ for $\psi \approx 1$. However, as was
already emphasized by Pringle (1992),
$\nu_1$ and $\nu_2$ are physically distinct quantities, so
$\psi$ in general may well differ from 1. This parameter may, in
fact, be expected to be less than 1 in a disk where a
centrifugally driven outflow is the dominant angular momentum
transport mechanism. In particular, if one uses the ``$\alpha$
prescription'' to represent the two effective
viscosities, then, while $\alpha_2$ (which likely corresponds to a real
viscous stress) is $\lesssim 1$,
$\alpha_1 \approx (|v_r|/C) (r/h)$ [where $h$ is the
disk scale height, $v_r$ is the radial inflow velocity,
$C$ is the speed of sound, and $\alpha_1$ now describes the
effective ($z,\phi$) viscosity] could potentially be $\gg 1$
(since $|v_r| \lesssim C$ in such a disk and $r/h \gg 1$;
see, e.g., Wardle \& K\"onigl 1993). If the disk in NGC 4258
drives a hydromagnetic wind, it may thus in principle be both
advection dominated and warped by a radiation-induced instability.

Another potentially relevant aspect of a disk-driven
hydromagnetic wind is related to the results of the global stability analysis
carried out by Maloney et al. (1996b), who generalized the local
treatment of Pringle (1996). They showed
that radiation-induced warping instabilities in AGNs
would be dominated by modes that peak near the outer radius of
the disk, identified as either the physical edge of the disk or
the radius where the disk becomes optically thin to the bulk of the
incident or reemitted flux. Since a hydromagnetic disk outflow
is expected to be dusty beyond the sublimation radius (see
KK94), the radial flux component at the disk surface
(which enters into the calculation of the torque on the disk)
would be attenuated by passage through the wind as a result of
dust scattering and absorption (as well as of electron scattering).
This would tend to shift the effective outer radius of the disk
inward, with the exact location determined by the density
profile and total mass outflow rate in the wind.

To summarize the discussion in this Appendix, a disk-driven
hydromagnetic wind may reconcile the radiation-induced warping
interpretation and the advection-dominated accretion model in a
source like NGC 4258. It may also be expected to affect the location
of a warp arising from this instability. A physical warp is a
necessary ingredient of the irradiated homogeneous-disk model of Neufeld \&
Maloney (1995), but, as discussed in \S\S 4 and 5, one could in
principle account for the megamaser emission from AGN disks even
if they were not warped. It is also worth reiterating our
conclusion in \S 2.1 that a radiatively heated maser disk in a
source like NGC 4258 is likely clumped even if warping (rather
than, say, wind ram pressure on clouds) is responsible for exposing the
masing gas to the central continuum.

%\clearpage

\clearpage

\centerline{\bf Figure Captions}

\figcaption {The minimum shielding column density of dusty gas
that is required for photoionized clouds to be molecular, as a
function of the reduced pressure $p/k = n_{\rm H} \, T$ of the
irradiated clouds. The two curves correspond to clouds that are
located at a distance of $1\ \pc$ and $0.25\ \pc$ from the
central continuum source in NGC 1068 ($L_{\rm
bol} \approx 4 \times 10^{44}\ {\rm ergs\ s^{-1}}$) and NGC 4258 ($L_{\rm
bol} \approx 10^{42}\ {\rm ergs\ s^{-1}}$), respectively. Fitting these
curves by an expression of the form ${\rm log}\ N_{\rm min} = c_1
+ c_2 x + c_3 x^2$, where $x = {\rm log}\ (p/k)$, one finds
$\{c_1,c_2,c_3\} = \{47.10,-3.29,0.10\}$ and
$\{14.46,1.99,-0.12\}$ for NGC 1068 and NGC 4258, respectively.
\label{fig1}}

\figcaption {Schematic diagram of a clumped
ring that orbits a black hole in an AGN. In an
attempt to mimic the inferred conditions in NGC 1068, the ring
plane is shown tilted at an angle of $45^{\circ}$ with respect
to the equatorial plane, whereas the radio jet axis is depicted
as being normal to that plane. For a similar reason, the ratio 
of the disk width to its inner radius was taken as $\Delta r/r_{\rm in} \approx
0.6$. The ring rotation axis is shown at an
inclination of $80^{\circ}$ with respect to the line of sight.
The gas in the ring is assumed to be dusty (consistent with the
fact that the dust sublimation radius lies interior to $r_{\rm
in}$) and to be composed of cold, dense clumps and possibly also
a warm, more tenuous interclump medium. In the case of NGC 1068
there is evidence for ionized gas that produces free-free radio emission
within $r_{\rm in}$. This emission (which may in part originate in the
inner rim of the ring) could be amplified into maser radiation
by clumps located near $\rin$. High-velocity maser features may
be associated with self-amplifying clumps situated along the ring midline. 
\label{fig2}}

\figcaption {Phase-space diagram in the shielding-column
vs. fractional-H$_2$O-abundance plane of the ``minimalist''
irradiated-ring model for NGC 1068. The dusty-gas column densities $N_{\rm
min}$, $N_{\rm max}$, and $N_{\rm cent}$ correspond,
respectively, to the minimum shielding that is required for a
cloud to be molecular (see Fig. 1), the maximum shielding that
allows the temperature of a cloud to exceed $\sim 250\ {\rm K}$
so that collisional pumping can operate (which, in turn,
determines the outer maser radius $r_{\rm out}$), and the
shielding of the central continuum by intervening clouds (see
eq. [20]). All these columns depend on the cloud density $\nc$
and were calculated for a cloud located at $r=1\ \pc$. In order
for the model to be self-consistent, the water abundance must be
such that $N_{\rm cent}$ for the given cloud density lies
between the corresponding values of
$N_{\rm min}$ and $N_{\rm max}$. To guarantee that the ring is
not dispersed by radiation pressure on the dusty gas, $N_{\rm
max}$ should not be much smaller than $N_{\rm rad}$ (eq. [21]).
\label{fig3}}

\figcaption {Fractional water abundance ({\em top} panel) and
gas temperature ({\em bottom} panel) as a function of distance
$d=10^{14}\, d_{14}\ {\rm cm}$ into a dusty, uniform-density slab that is
irradiated by the nuclear continuum of NGC 1068 (assumed to be unattenuated
before it reaches the slab) at a distance of $1\ \pc$ from the
center. The curves are labeled by the value of the slab density
$n_{\rm H} = 10^9\, n_9\ \cmm$ and terminate at the point where
the temperature declines to $250\ {\rm K}$.
\label{fig4}}

\figcaption {Same as Fig. 3, but for the continuum-source parameters of
NGC 4258 and $r=0.25\ \pc$. Because of the comparatively low
luminosity of this source, the shielding column $N_{\rm rad}$
(eq. [21]) is not a relevant constraint in this case.
\label{fig5}}

\figcaption {Same as Fig. 4, but for the continuum-source parameters of
NGC 4258 and $r=0.25\ \pc$.
\label{fig6}}

\figcaption {($a$) Schematic diagram illustrating the expected regions of
efficient maser emission from clouds embedded in accretion
disk-driven hydromagnetic winds that are viewed orthogonally to
the rotation axis.  The {\em light dashed} curve indicates the
dust sublimation surface in the wind, while
the shading demarcates the masing zones: along the sublimation surface for
high-density winds, and along the disk surface for lower-density winds.
The {\em heavy dashed} curves show poloidal projections of
representative cloud 
trajectories.  Lighter clouds (of mass $M_{\rm c} \lesssim M_{\rm
c,\, max}$; see eq. [24]) tend to move along wind streamlines and
rise to high altitudes, whereas heavier clouds ($M_{\rm c}
\gtrsim  M_{\rm c,\, max}$) tend to move across wind
streamlines and remain close to the disk surface.  The maser
features in a source like NGC 4258 are evidently best
modeled by a comparatively low-density wind that gives rise to
a narrow efficient-masing strip near the disk surface. ($b$)
Line-of-sight velocity as a function
of (cylindrical) radius for clouds moving in a disk-driven
wind. The wind is assumed to be radially self-similar and
is described by model A1 in KK94 (with a total
mass outflow rate of $0.03 \ M_{\odot}/{\rm yr}$ and a central
mass $M_{\rm bh} = 3.5 \times 10^7 \ M_{\odot}$). The {\em solid}
and {\em dashed} curves represent clouds with
masses $M_{\rm c} \approx 0.5 \ M_{\rm c,\, max}$ and $M_{\rm c} \approx
11 \ M_{\rm c,\, max}$, respectively. The latter curve adequately
describes the nearly Keplerian velocity field of the maser
features in a source like NGC 4258.
\label{fig7}}

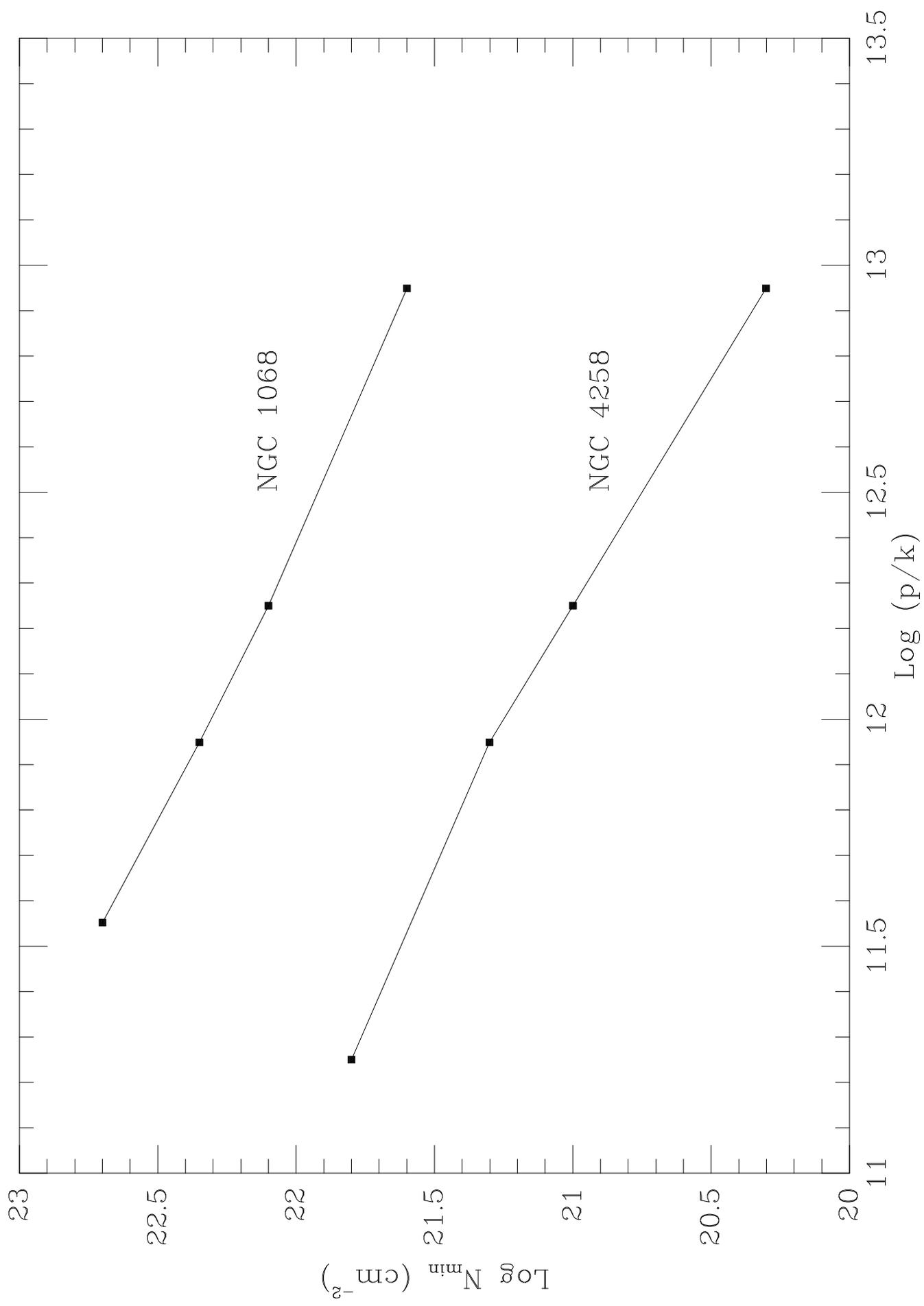
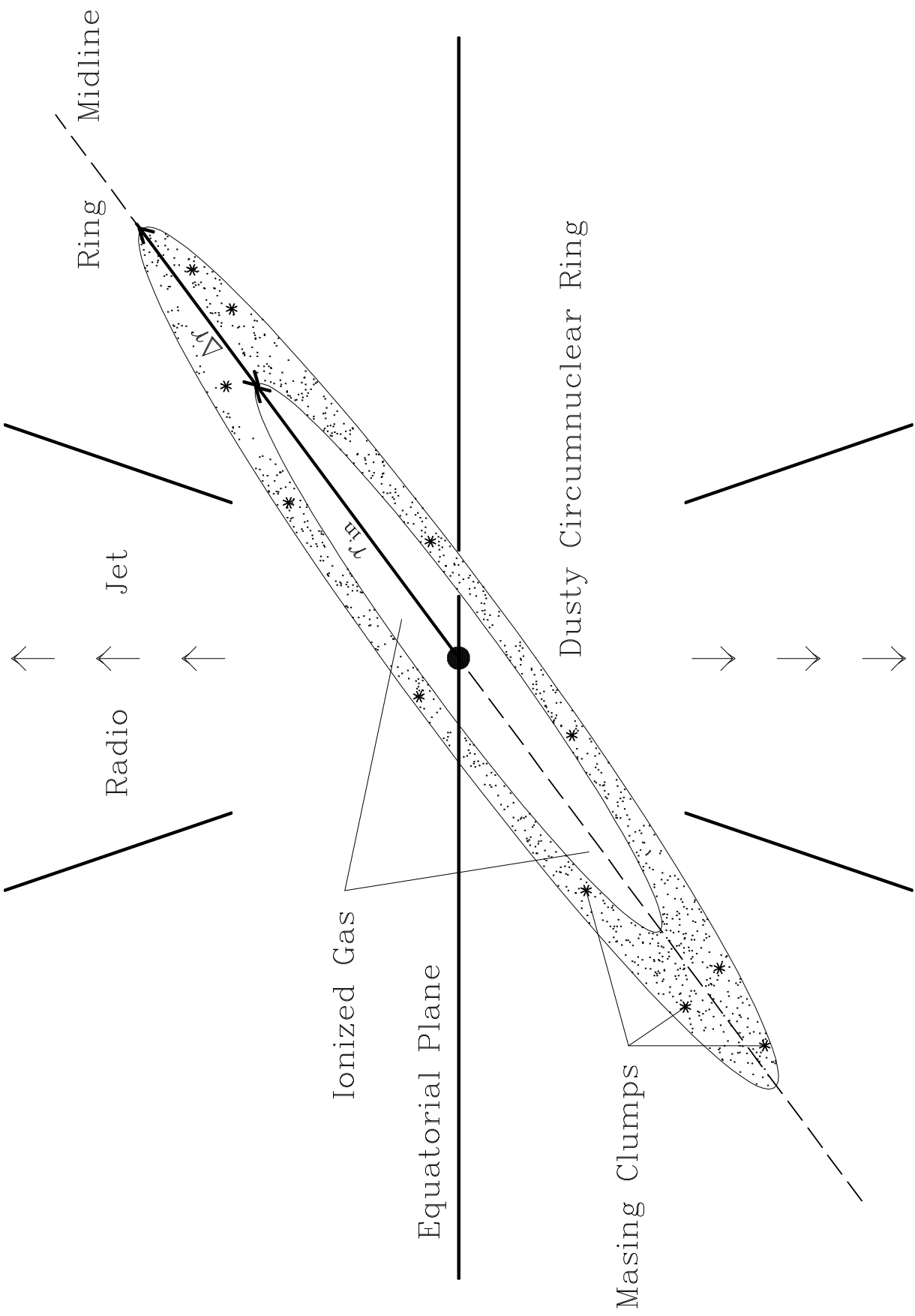
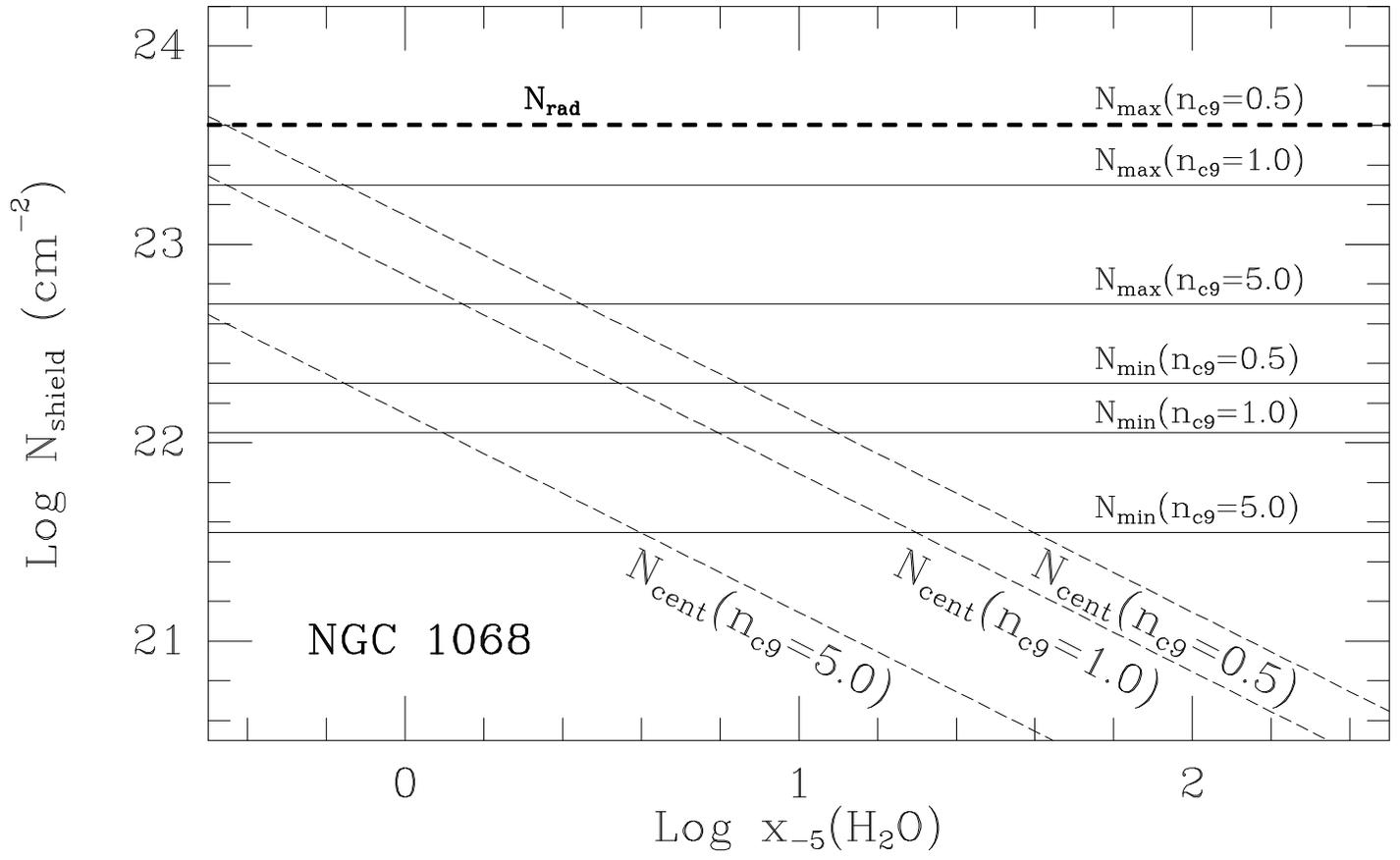
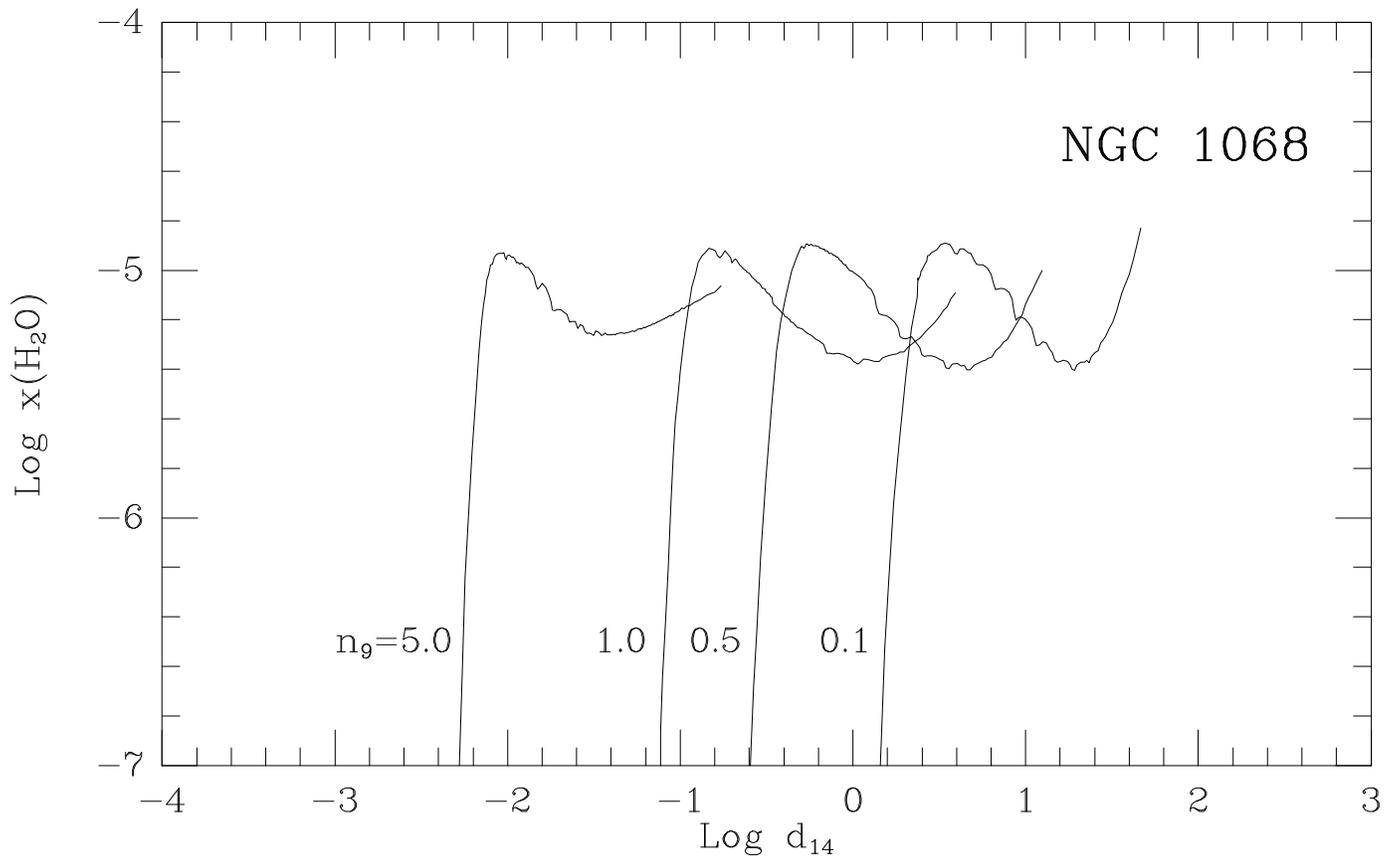
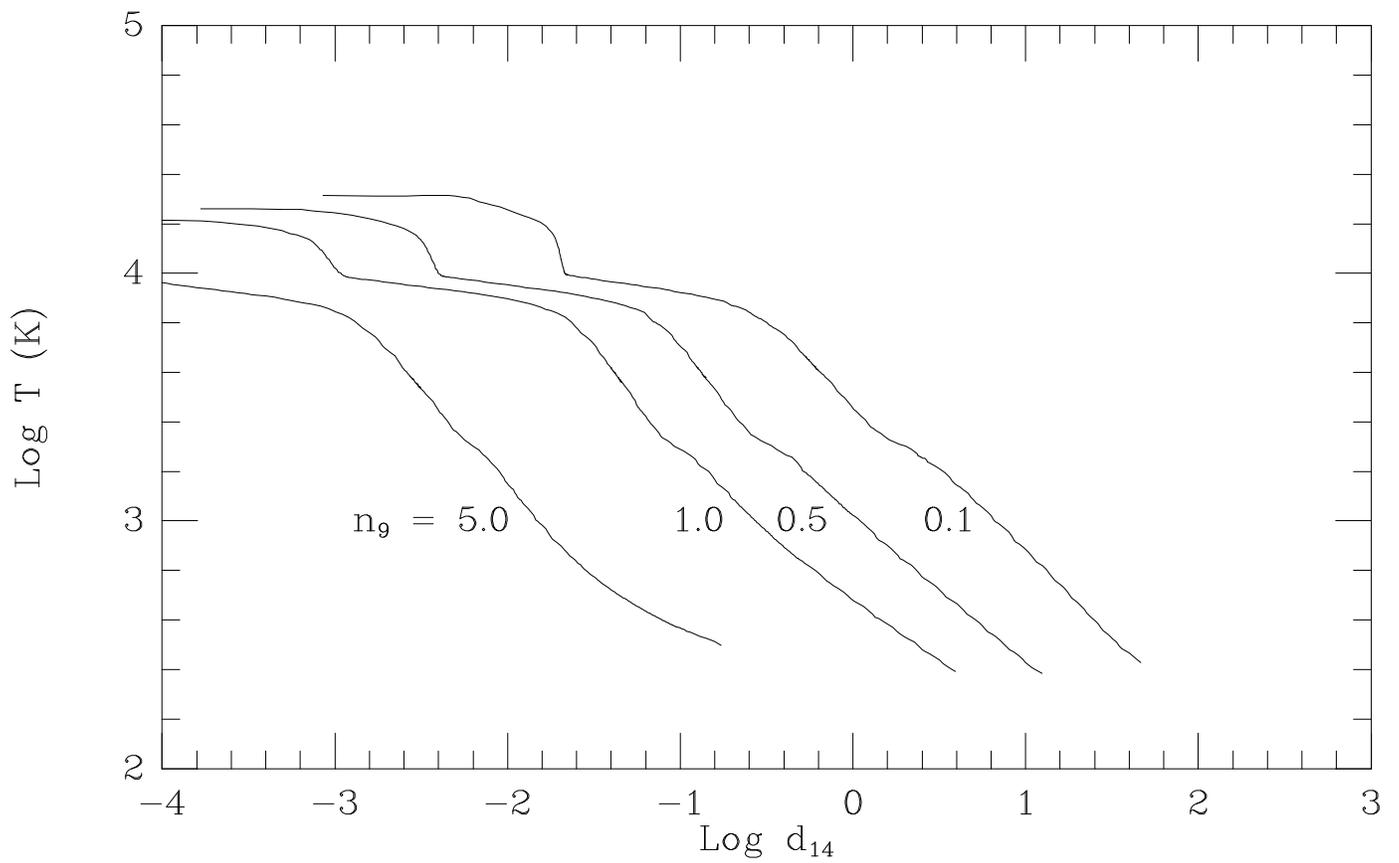
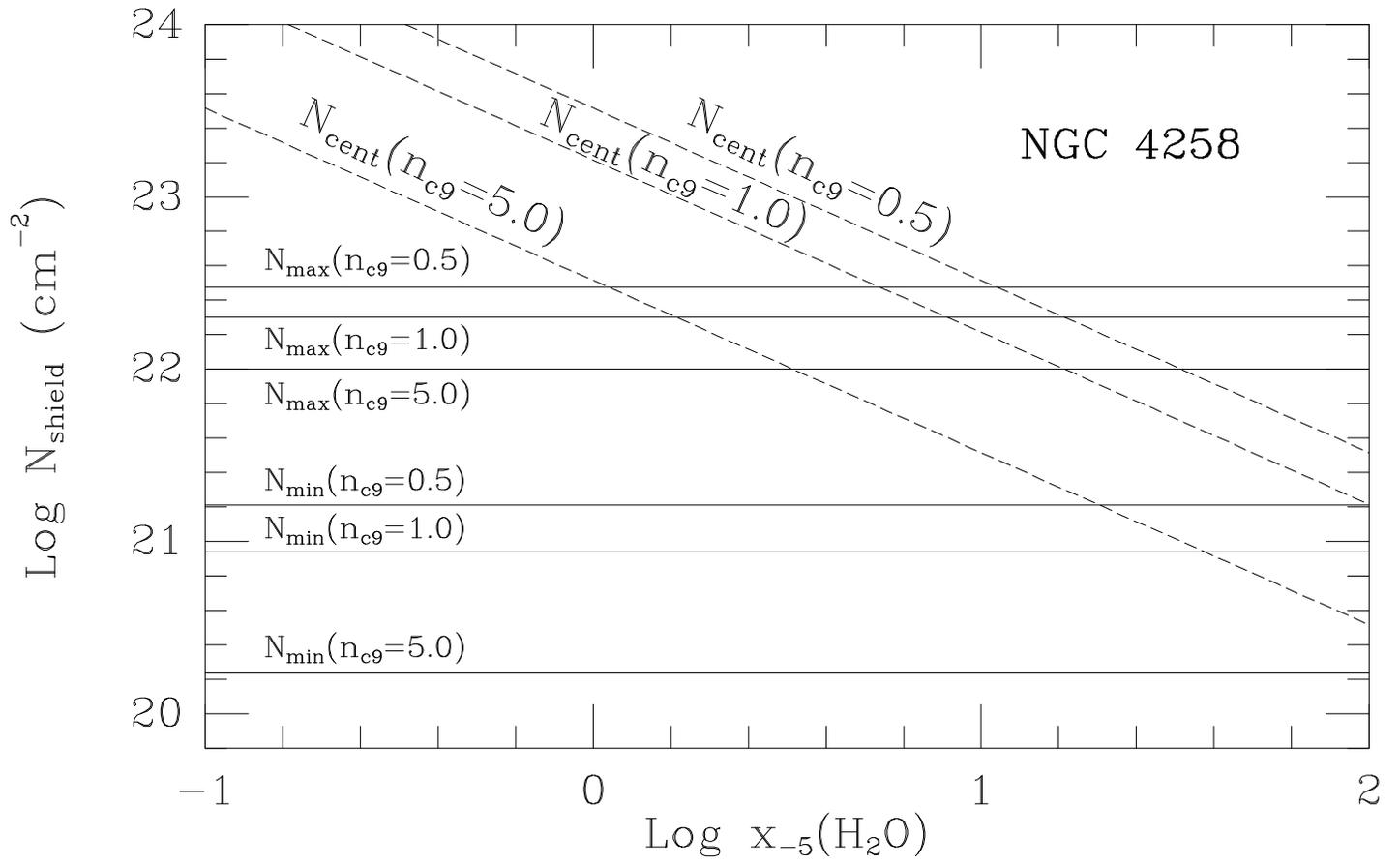
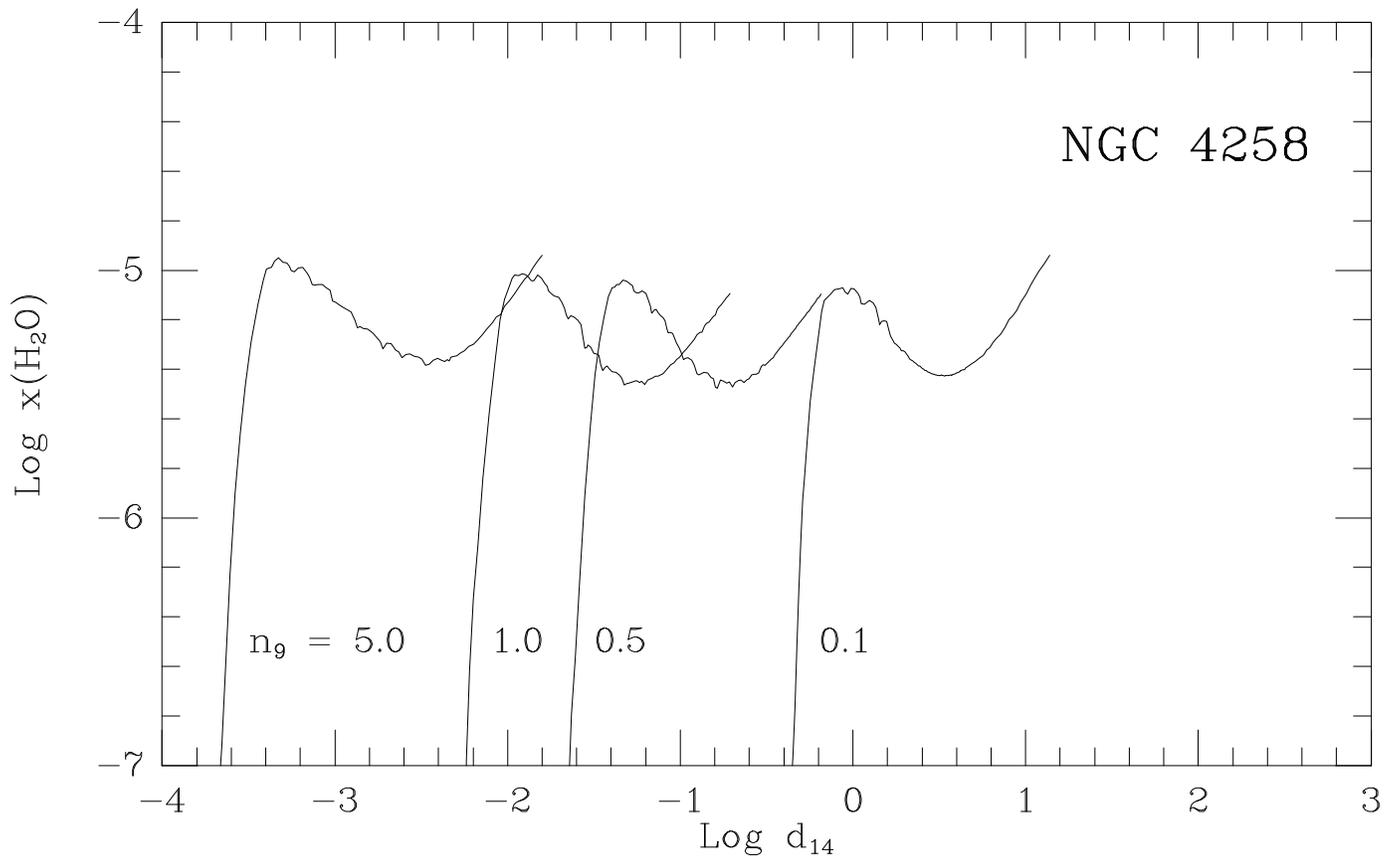
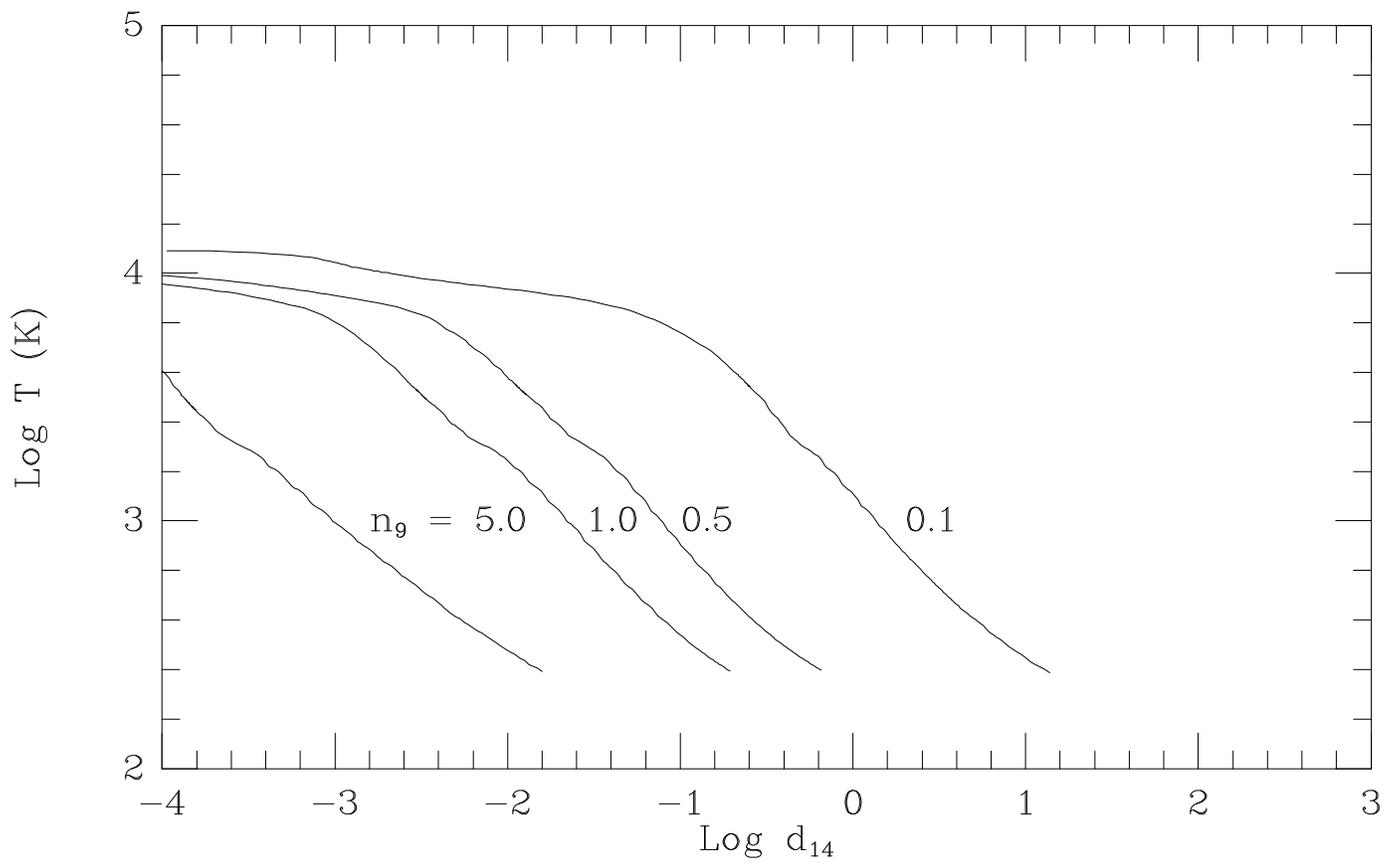
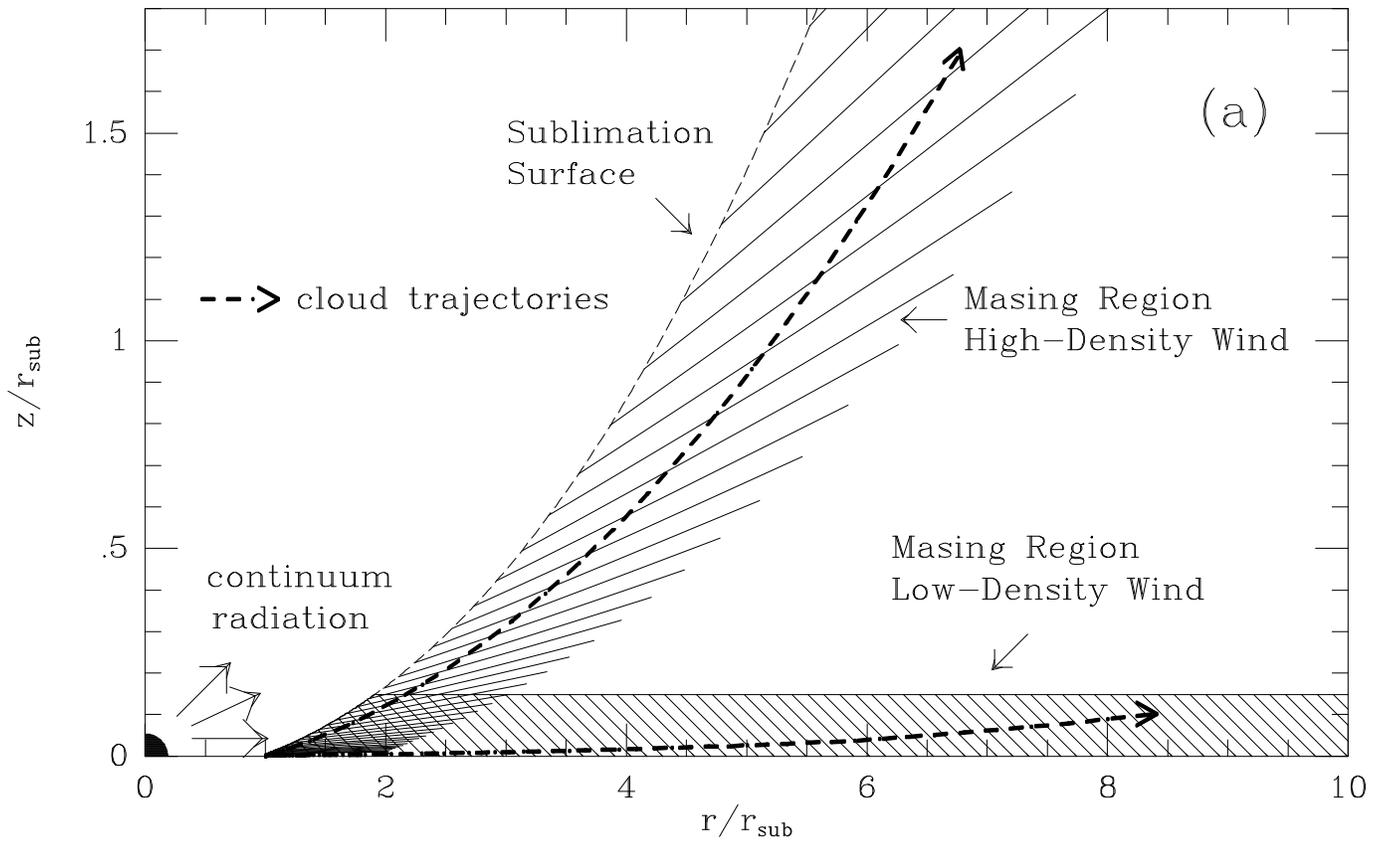
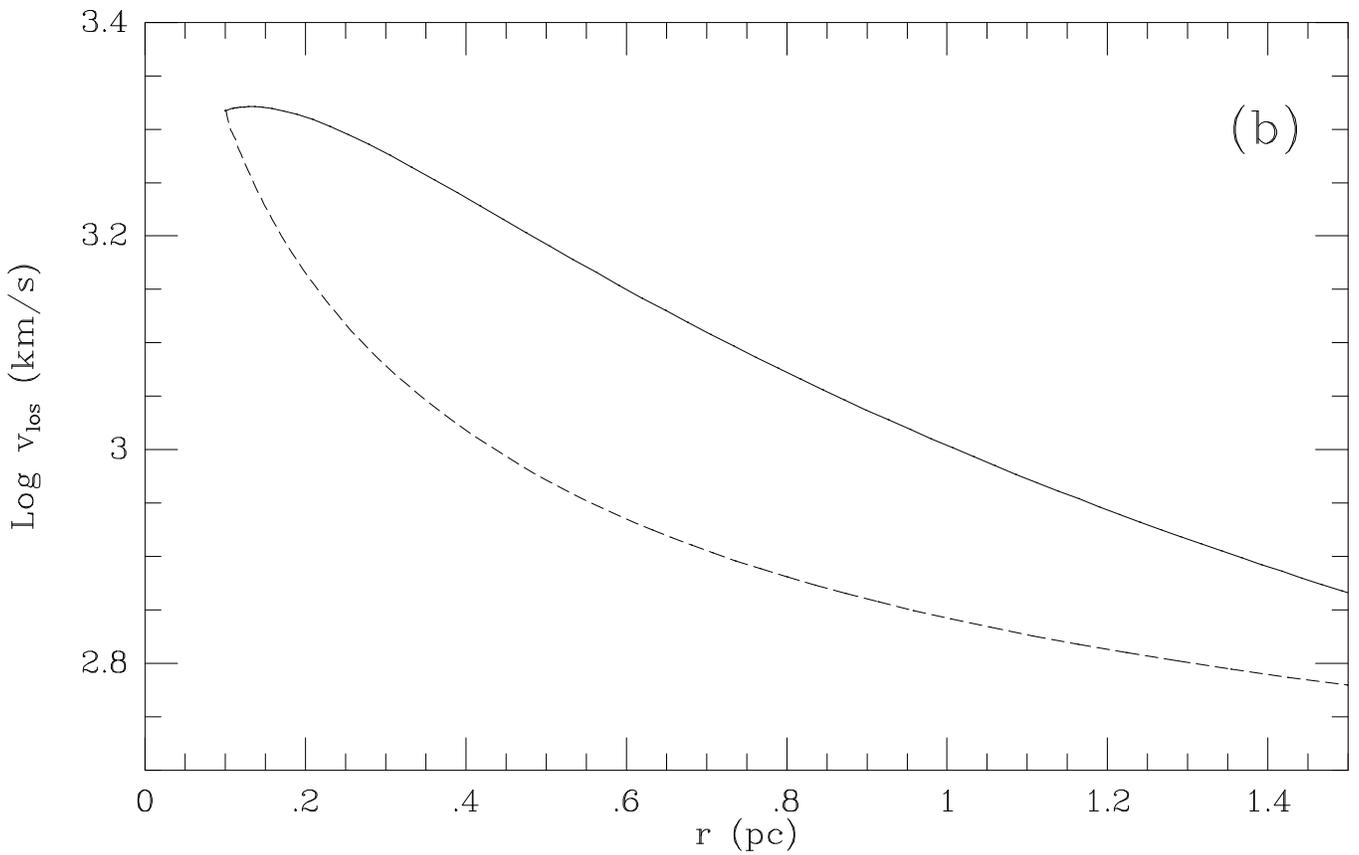
\end{document}